\documentclass[11pt]{article}
\usepackage[utf8]{inputenc}
\usepackage{cmap}
\usepackage[T1]{fontenc}
\usepackage{amsmath,amssymb}
\usepackage{graphicx}
\usepackage{float}
\graphicspath{{figures/}{./}}
\usepackage{booktabs,array,tabularx}
\usepackage{xltabular}
\usepackage[margin=1in]{geometry}
\usepackage{xcolor}
\definecolor{boxbg}{RGB}{244,248,251}
\definecolor{boxln}{RGB}{150,170,190}
\usepackage{mdframed}
\usepackage[hidelinks]{hyperref}
\usepackage{caption}
\usepackage{parskip}
\title{Three Failures of Pain Location: Why Its Diagnostic Utility Is Three Quantities, Not One}
\author{Adam Y. Shavit\\ \small Hunter College and the Graduate Center, CUNY \\ \small ORCID: \href{https://orcid.org/0009-0008-1235-0995}{0009-0008-1235-0995}}
\date{}
\begin{document}
\maketitle
\begin{abstract}Patient-reported pain location is diagnostically decisive for some presentations and nearly uninformative for others. The prevailing account treats this as a single gradient of diagnostic utility governed by anatomical complexity. That explanation conflates three epistemically distinct failures of localization, each with its own mathematical structure, matching instrument, and public-health consequence. In \emph{anatomical multiplexing} (a), many structures can produce the same felt location, so the source cannot be recovered from location alone: a non-identifiable inverse problem. In \emph{delocalized amplification} (b) — clinically, central sensitization or nociplastic pain, meaning pain attributed to altered nociceptive processing without a local lesion sufficient to explain it — the dominant generative class shifts toward a centrally maintained pain-behaviour \emph{pattern}, and mixed peripheral and central contributions can coexist clinically: a change of generative model. In \emph{group-conditional presentation shift} (c), location is hypothesized to shift in a systematic, person-dependent way, as in atypical presentations: a conditional shift in the presentation likelihood whose direct evidence is still open, distinct from a well-evidenced finding about what happens after diagnosis — in the care pathway, not in perception — that better detection alone does not improve outcomes when treatment uptake lags. Each of the three needs a different remedy, and a single principle unifies them: they are one Bayesian inference problem failing at different \emph{nodes} — different points in one generative model: the likelihood, the model class, and group- or context-dependence in that same likelihood. A fourth node is the report itself, the \textbf{reporting node}, where expectation and the measuring instrument move where pain is \emph{said} to be. Decision and treatment follow that model; they are not inside it. Keeping them out stops a treatment disparity from being read as a perceptual one. The formal development of all four nodes — inverse-problem, information-theoretic, decision-theoretic, neural-field, and a spatial Bayesian model of the report — is in a companion paper (Shavit, 2026); this paper states what each model shows and what follows clinically.\par \textbf{The reframing is new, and it changes what the evidence supports.} Re-examination finds that the published “high-utility” accuracy band leans on overstated specificity (Lipton et al., 2003; Bruyninckx et al., 2008; Devillé et al., 2000), so the gradient is real but flatter than drawn. Observation over time cannot reduce the recoverable information about a fixed cause, but that is a property of an idealization and not a guarantee that assessing a patient repeatedly beats assessing them once. And the well-evidenced finding that better detection alone does not improve outcomes when treatment uptake lags is about the care pathway, not about perception — a distinction the three-way split makes visible and a single “diagnostic utility” number hides. This paper organizes the failures along a “why-location-fails” axis, distinct from the nociceptive/neuropathic/nociplastic taxonomy (Kosek et al., 2016), and sets out the study that would test its most discriminating open prediction.\end{abstract}

\par\medskip

\begin{mdframed}[backgroundcolor=boxbg,linecolor=boxln,linewidth=0.4pt]\small \textbf{Disclaimer.} This is a conceptual and methodological Perspective. The models are illustrative
existence proofs, not validated clinical decision tools; the diagnostic figures are drawn from
the cited literature to make structural points, not to guide individual care. Nothing here is
medical advice or a substitute for professional clinical judgment.\end{mdframed}

\emph{Preprint on arXiv (q-bio.NC + stat.AP). APA style, author-year in-text citations. Model code, the verification trail, and the preregistration are archived in the public supplement (Open Science Framework, \url{https://doi.org/10.17605/OSF.IO/ZFHKQ}). Simulation results are illustrative existence proofs (see Limitations), and the literature synthesis is targeted pending a registered systematic search.}

\textbf{The models are developed in a companion paper}, "One Inference, Four Nodes: Formal Models of Why Pain Location Fails" (Shavit, 2026), which carries the formal treatment of all four nodes. This paper states what each model shows and what follows clinically; it does not derive them. Pointers of the form "companion §7" name a section of that paper.

\section{Key Points}

\begin{itemize}\setlength{\itemsep}{2pt}\item \textbf{One diagnostic-information number, three different failures that drive it down.} The same low value of that number can arise from three failures with different mathematics, different matching instruments, and different remedies, so the number alone cannot identify \emph{why} location is uninformative, or which further measurement would discriminate best.\item \textbf{Where many structures share one location, the problem is identifiability, not resolution.} For \emph{deep and visceral} sources, where competing structures produce nearly indistinguishable location distributions, refining the same location measurement cannot separate them; provocation adds source-discriminating measurements instead. (Localization on the skin is precise — around 5 mm for mechanically evoked pain — so this is a claim about depth, not about pain reports in general.) Single manual sacroiliac tests are uninformative in isolation, while three or more positive tests pool to a positive likelihood ratio (LR+) of 3.2 (95\% CI 2.3–4.4), clearing the review's informativeness threshold of 2 — roughly a threefold multiplication of the pre-test odds.\item \textbf{Where pain is centrally amplified, the model predicts that the spatial pain map is non-stationary over time.} If extent and location drift across days, a single assessment cannot measure the feature that defines the mechanism, and the estimates examined here rest on single assessments (Balasch-Bernat et al., 2022; Fernández-de-las-Peñas et al., 2022). Non-stationarity is the paper's least-supported prediction and the one it most wants tested.\item \textbf{Whether felt or reported pain location itself shifts by patient group is not yet directly tested; a related but separate finding, about what happens after diagnosis, is well established.} Detection improved while outcomes did not in two settings — a high-sensitivity troponin rollout and routine early imaging for low-back pain — but those trials tested diagnostic reclassification and treatment uptake, not location displacement. The model reads them as a downstream care-pathway corollary: better detection alone does not improve outcomes when the bottleneck is treatment, not detection.\item \textbf{A fourth failure happens after the other three, at the report itself.} Pain that was felt correctly can still be misplaced when it is reported, and that stage is modelled explicitly in the companion paper.\end{itemize}

\section{Introduction}

Cross your wrists so the palms turn outward, interlace the fingers, then rotate the clasped hands down, inward, and up under your chin. Now have someone point at one of your fingers and ask you to lift it. Most people cannot. They hesitate, lift an adjacent finger, or lift the matching finger on the wrong hand. Have that person \emph{touch} the finger instead and the confusion disappears. Nothing about the hand changed between the two attempts. Only the way the target was specified changed — named versus touched (Figure 1).

\begin{figure}[H]\centering
\includegraphics[width=\linewidth]{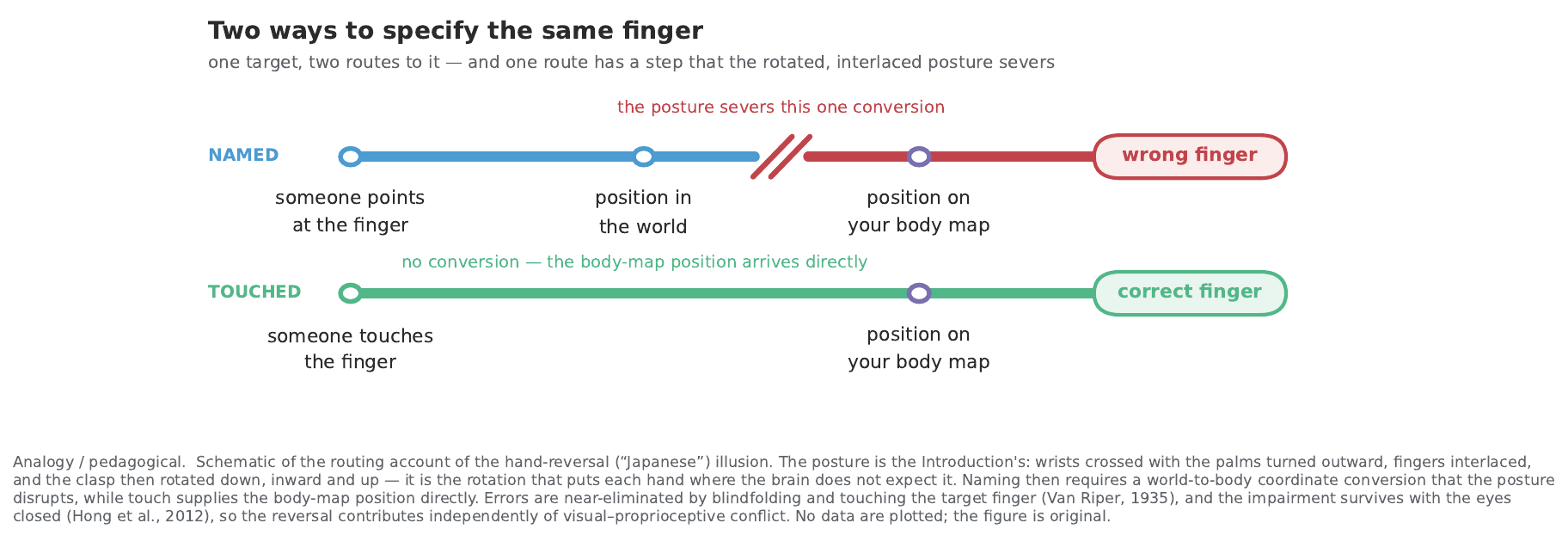}
\caption{Figure 1. Two ways to specify the same finger. Naming requires a conversion from a position in the world to a position on your own body map, and the crossed, interlaced posture breaks that conversion — so the named finger comes back wrong. Touch delivers the body-map position directly, with no conversion to break, and the same finger is identified correctly. Nothing about the hand changes between the two attempts; only the route to it does. Errors are near-eliminated by blindfolding and touching the target finger (Van Riper, 1935), and the impairment survives with the eyes closed (Hong et al., 2012), so the reversal contributes independently of visual–proprioceptive conflict. [Analogy / pedagogical — the routing account drawn; no data are plotted.]}
\end{figure}

The effect has a name: the hand-reversal, or “Japanese”, illusion (Van Riper, 1935). The crossing causes the effect. It puts each hand on the side the brain does not expect, so the usual correspondence between a finger \emph{seen} and a finger \emph{felt} no longer holds. Pointing gives the finger's position in the world, and to act on it you must convert that into a position on your own body — the conversion that the crossed posture breaks. Touching delivers the body-map position directly, so no conversion is needed. The errors are virtually eliminated by blindfolding the participant and touching the target finger rather than naming it (Van Riper, 1935). Hong et al. (2012) measured reaction time, not error rate, and found that the impairment survives with the eyes closed. Visual–proprioceptive conflict therefore does not account for it alone: the reversal of left and right hands in external space contributes independently. That is a remapping failure. The felt body is not read off the flesh; it is a \emph{map}, and reaching it from outside depends on a calibration that can be knocked out of alignment.

Nociception — the nervous system's detection and signalling of tissue damage, which precedes the felt experience of pain and is not the same thing as it — begins in tissue. But the pain, and the \emph{place} it is felt, is constructed centrally and then assigned to a location on the body. A nociceptor is not a switch that turns on a sensation already labelled with its address. The address is computed. Anything computed can be computed wrongly, and the mirror box is the standing illustration. Show a patient a reflection of the intact hand where the amputated one should be, and pain in a limb that no longer exists can ease — first reported with the mirror box in a small case series, where mirror-assisted opening relieved painful phantom clenching in four of five affected patients (Ramachandran \& Rogers-Ramachandran, 1996), and since tested in a randomized controlled trial of mirror therapy for phantom-limb pain (Chan et al., 2007). The result is consistent with visual feedback recalibrating the map: manipulating it can change the pain even when the state of the tissue has not changed — so pain is not a simple readout of tissue.

Those cases show why \emph{where it hurts} is diagnostic at all, and why it is not always. Clinicians know that a unilateral throbbing headache suggests migraine, that dermatomal leg pain suggests radiculopathy, and that exertional substernal pressure suggests angina — impressions that emerge largely from \emph{where} a symptom is felt. Those inferences work when the map is well calibrated for the structure in question. The same inferences fail for diffuse low-back pain, widespread pain, and undifferentiated chest pain, which is conventionally explained as a single gradient of diagnostic utility set by anatomical complexity.

This single-gradient framing conceals more than it reveals. If location is a computed, calibrated assignment and not a direct readout, then “low utility” is not one property of a body region. It is whatever has gone wrong with the mapping. A mapping can fail in more than one way. “Low utility” therefore labels at least three distinct underlying problems that share only a surface feature: location does not lead cleanly to a diagnosis. Once separated, they demand different instruments, imply different limits, and carry different implications for equity and for the appropriate use of imaging.

Throughout, \emph{diagnostic utility of location} carries one specific sense, chosen to tie the three fields' measures together: how much knowing where the pain is felt reduces uncertainty about what is causing it. The felt location is the \emph{mapped} one: what the calibration produced, not the coordinates of the tissue itself. What reaches the clinician is one step further on — the patient's \emph{report} of that felt location, which the fourth failure below can displace again. Formally, that reduction in uncertainty is the mutual information between cause and felt location, written $I(s; \ell )$ for cause $s$ and location $\ell $ (companion §5). Sensitivity, specificity and likelihood ratios are threshold-specific summaries of that same joint distribution, not direct readouts of it: a likelihood ratio at a chosen threshold does not recover the mutual information without the prevalence and the full joint distribution (companion §5). The information becomes a decision only through the clinical loss of acting on it (companion §6). Mutual information is the threshold-free information content; the accuracy statistics are its familiar clinical summaries; the loss is what turns information into \emph{utility}. Each of the three failures below drives this quantity down through a different one of them.

Every model, figure, and sweep reported here regenerates from the public archive at \url{https://doi.org/10.17605/OSF.IO/ZFHKQ}, which also holds the preregistration and the model-testing registry (see Data and Code Availability).

\section{A Reframing: Three Epistemically Distinct Failures}

Each of the three is a different way the map-and-calibrate step fails: the map is many-to-one, the map goes unstable, or the map is read out with a systematic offset.

In \emph{anatomical multiplexing} (a), several structures share one location — the lumbar spine is the standard case — which makes recovering the source a non-identifiable inverse problem. In \emph{delocalized amplification} (b), the dominant generative class shifts toward a centrally maintained pain-behaviour pattern rather than a peripheral generator, as in fibromyalgia and nociplastic pain (Kosek et al., 2016; Nijs et al., 2021); this is a change of generative model. In \emph{group-conditional presentation shift} (c), location is hypothesized to shift in a systematic, person-dependent way, as in atypical myocardial infarction in women; this is a conditional shift in the presentation likelihood whose direct evidence is, unlike (a) and (b), still open (see below). These are not three points on one axis. The labels name failure components, not mutually exclusive patient types: more than one can operate in the same case, and the empirical problem is to estimate which of them costs information and how much. The reframing, not the mechanisms themselves, is the contribution.

\begin{figure}[H]\centering
\includegraphics[width=\linewidth]{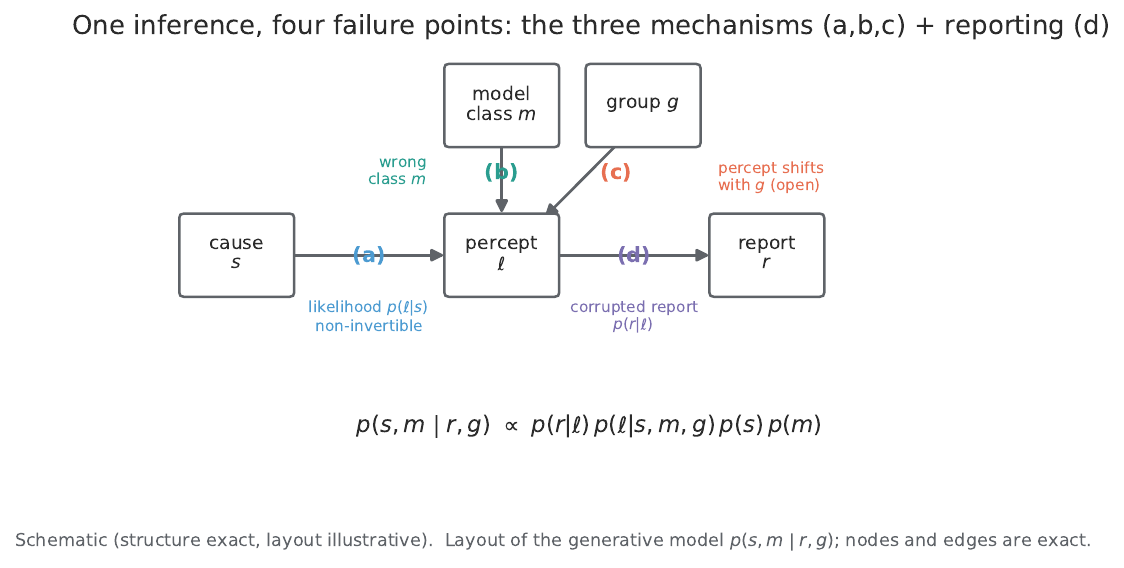}
\caption{Figure 2. One inference, four failure points, one decision layer downstream. The three mechanisms (a–c) and the reporting node (d) are distinct failures of a single generative model — probabilities only. Follow the arrows left to right: cause s becomes percept $\ell $ becomes report r, and each labelled edge is one place that chain breaks — (a) at the likelihood, (b) at the model class chosen, (c) at group- or context-dependence in that same presentation likelihood (open; not yet directly tested), (d) at the report itself. The factorization drawn here says, in words: infer the cause and the model class by combining how a felt location would generate the report, how a cause and model class would generate that felt location, and the prior plausibility of each cause and model. Decision, treatment, and outcome are drawn as a separate downstream layer reading the posterior out, not as additional nodes inside it — a loss is not a probability. [Schematic — node/edge structure exact, layout illustrative.]}
\end{figure}

\begin{figure}[H]\centering
\includegraphics[width=\linewidth]{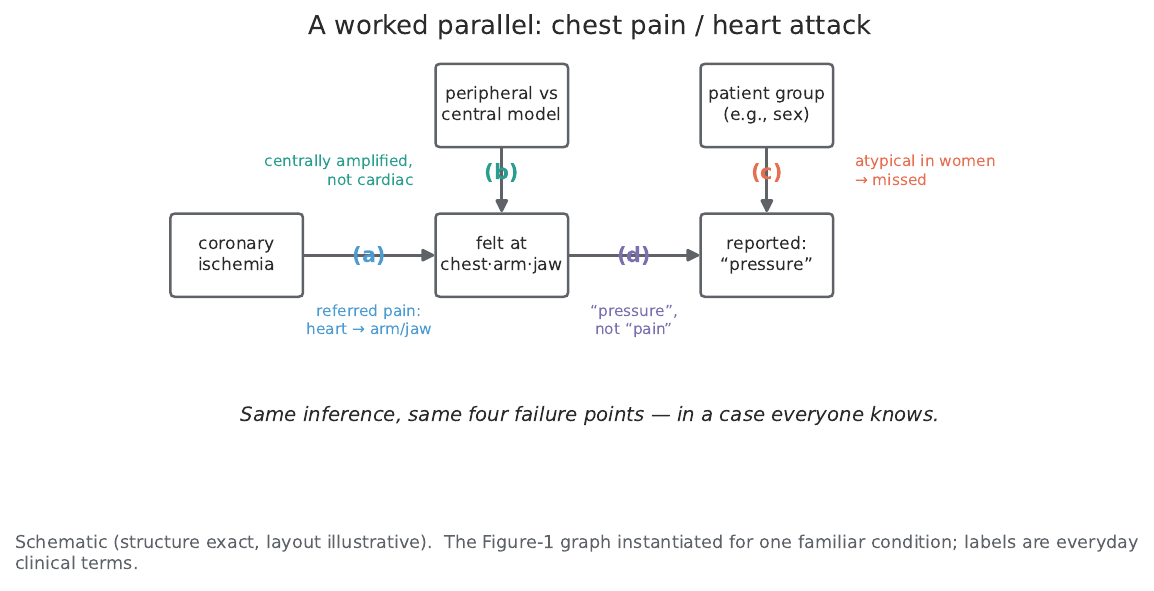}
\caption{Figure 2b. The same graph in a case everyone knows: chest pain / heart attack. Each abstract node and each of the four failure points carries an everyday clinical label — (a) referred pain (heart felt in the arm/jaw); (b) a model-class switch — recurrent chest pain that is centrally amplified (nociplastic, centrally sensitized) rather than cardiac, so a peripheral workup keeps returning negative; (c) atypical presentation in women, more often missed; (d) the report itself distorted (“pressure,” not “pain”). The instantiation is illustrative, not exact: (a) is a displaced projection rather than many-to-one ambiguity, (c) is a detection fact rather than the open location shift, and (d) is a difference of quality rather than of place — the three loosenings are named in “A Reframing”. [Schematic — the Figure-2 graph instantiated for one familiar condition.]}
\end{figure}

\subsection{Reading across three fields: a shared vocabulary}

This analysis is written for three audiences — sensory science, clinical medicine, and applied mathematics — and its central risk is that a term meaning one thing to one field means something else to another. The tables below therefore make the translations explicit.

\emph{The same object, three vocabularies.} The core question is how informative \emph{where a symptom is felt} is about \emph{what is wrong}. For sensory science that is the fidelity of spatial localization of a percept; for medicine, the diagnostic value of symptom location; for mathematics, the invertibility of $p(observation | cause)$. The three mechanisms translate as follows.

\emph{Table 1. One concept, three vocabularies.}

{\footnotesize\begin{xltabular}{\textwidth}{>{\raggedright\arraybackslash}X>{\raggedright\arraybackslash}X>{\raggedright\arraybackslash}X>{\raggedright\arraybackslash}X}
\toprule
Concept & Sensory science & Medicine & Mathematics \\
\midrule
\endhead
Anatomical multiplexing & convergent projection; poor spatial resolution & non-specific pain; many generators, one region & non-identifiable inverse problem (a rank-deficient referral matrix $R$, mapping sources to felt locations, for a continuous source; a categorical source needs a restricted criterion instead — companion §4) \\
Delocalized amplification & delocalized, gain-amplified percept & central sensitization / nociplastic pain & a threshold-driven regime change (bifurcation) \emph{within} one dynamical model, and — separately — model selection \emph{between} a peripheral and a centrally amplified account \\
Displacement & population/observer differences in report & atypical presentation; demographic bias & group-conditional presentation shift (open); decision-threshold problem downstream \\
“Dynamics” & temporal structure of perception & longitudinal symptom tracking & repeated channel use $\to $ information gain \\
\bottomrule
\end{xltabular}}

\emph{Terms that mean different things in each field (a cross-audience map).} Several words mean different things to the three fields, and a silent mismatch is the main way this account could be misread. The table gives, for each easily-mistaken term, what each audience is likely to assume and the sense used here; each is also defined at first use. Two \emph{terms} carry two meanings even inside this paper's own text, and are flagged where they do.

\emph{Table 2. Terms that mean different things in each of the three fields.}

{\footnotesize\begin{xltabular}{\textwidth}{>{\raggedright\arraybackslash}X>{\raggedright\arraybackslash}X>{\raggedright\arraybackslash}X>{\raggedright\arraybackslash}X>{\raggedright\arraybackslash}X}
\toprule
Term & In psychology & In medicine & In applied mathematics & In \emph{this paper} \\
\midrule
\endhead
Sensitivity & perceptual sensitivity (d$'$) & a test's true-positive rate & $\partial $output/$\partial $input (sensitivity analysis) & as in medicine \\
Sensitization & learning-theory sensitization (response grows with repetition) & central sensitization (clinical construct) & — & a neural \emph{gain} increase; not learning-sensitization, not sensitivity \\
Specificity & specific vs. general & a test's true-negative rate & uniqueness / identifiability & as in medicine; not mathematical uniqueness \\
Resolution & spatial acuity (just-noticeable difference, JND) & symptom resolution (recovery) & inverse-problem resolution (matrix rank) & as in applied mathematics; not symptom recovery \\
Prior & a Bayesian prior belief & prior history & a prior distribution & as in applied mathematics; reads clinically as pre-test probability \\
Posterior & a Bayesian posterior & anatomical (toward the back) & a posterior distribution & as in applied mathematics; except “posterior insula,” which is anatomical \textbf{(!)} \\
Bias & response bias / criterion (signal detection theory, SDT) & systematic error; inequity & estimator bias, or an offset term & two senses, both flagged \textbf{(!)}; displacement = a person-dependent shift, Model E's \emph{b} = a fixed offset \\
Referral (\emph{R}) & a clinical referral & referred pain, \emph{or} a clinical referral & the matrix mapping sources $\to $ felt location (lead-field) & the referred-\emph{pain} matrix \emph{R}; not sending to a specialist \\
Bifurcation / “critical” & dynamical-systems term & “critical” = critically ill & a qualitative regime change at a threshold & a tipping point (focal $\to $ widespread); “critical” names the threshold, not severity \\
\bottomrule
\end{xltabular}}

A fuller version — including \emph{gain}, \emph{model class}, \emph{capture}, \emph{forced fusion}, and \emph{mutual information}, plus a figure-by-figure cross-audience reading — is in the three-audience reading guide in the supplement. Plain definitions of terms one audience may \emph{not know} (as opposed to the shared words above) are collected in the \textbf{Glossary (Appendix B)}.

Each mechanism below opens with an \textbf{Intuition} box: the idea in plain terms that all three fields share, before the field-specific formalism.

\emph{What each audience gains.} For sensory science, pain becomes a natural experiment in spatial localization under altered central gain. The bifurcation model (companion §7.1) yields a falsifiable directional prediction: under five stated sufficient conditions, finite-window migration \emph{declines} as central gain carries the system onto the high-barrier side of the transition. For medicine, each mechanism implies a different \emph{next measurement} to try: a structured interview and provocation for multiplexing, longitudinal tracking for amplification, group-specific criteria plus attention to the treatment gap for displacement. Naming the mechanism clarifies one reason further static imaging may fail to resolve source ambiguity, and points to the kinds of measurement that could; it is a research programme rather than a bedside rule, since none of these models is fitted. It also locates diagnostic inequity in a step that can be changed: the criteria applied and the treatment that follows. For mathematics, it is a live application in which inverse-problem, information-theoretic, decision-theoretic, and dynamical-systems tools each yield a clean structural result and a testable clinical prediction. The unifying Bayesian graph (Figure 2; companion §3) is the single diagram all three fields can point to at once. The same diagram maps onto a case everyone knows: chest pain and a possible heart attack (Figure 2b). It has four failure points. In referred pain, the heart's signal is felt in the arm or jaw. In a model-class switch, the chest pain is centrally amplified rather than cardiac in origin, so a peripheral work-up keeps returning negative. The presentation in women is missed more often than the presentation in men. And at the reporting stage, the patient says “pressure” rather than “pain”. That mapping is illustrative rather than exact, and it loosens in three places worth naming. Referred projection is the forward map sending one source to a displaced site, not the many-to-one ambiguity that defines multiplexing. Being missed more often is a downstream detection fact, not evidence that felt location differs by group, which stays the open hypothesis. And “pressure” for “pain” is a difference of quality rather than of place, so it sits beside the spatial reporting model rather than inside it.

\section{The Three Failure Modes}

Each mechanism gets the same treatment below, in the same order: what fails, the everyday intuition, the evidence for and against, and the instrument proposed to isolate it. The three are presented together because the claim is comparative. They differ not because each is individually novel, but because they respond to different instruments and are routinely pooled into one number anyway.

\subsection{(a) Anatomical multiplexing}

Disc, facet joint, sacroiliac joint, muscle and ligament can all produce pain in the same location. Inferring the source is then a many-to-one inverse problem: several distinct causes give the same felt location, so even a perfect reading of \emph{where} it hurts cannot separate them. The mathematics of this failure is general — any set of causes with indistinguishable observations produces it — but it matters most for \emph{deep and visceral} structures, because of their convergent projection. Localization on the skin is precise. On hairy skin of the hand dorsum, the threshold at which people are 75\% correct about where they were touched is 5.1 mm for mechanically evoked pain, 8.6 mm for heat pain and 9.0 mm for non-painful touch (Schlereth et al., 2001). Those thresholds fall within a fine whole-body acuity map (Mancini et al., 2014). They are lower than the 10–20 mm often quoted, and deliberately so. They come from a forced-choice task rather than from pointing, which removes the roughly 15 mm error the pointing movement itself contributes. The fingertip is finer again, at 1.3 mm — roughly four times finer than the 5.1 mm for mechanically evoked pain on the hand dorsum. The ambiguity is therefore not a property of the somatosensory map in general. It belongs to the convergent projection from deep and visceral generators, whose sparse, convergent innervation leaves them intrinsically diffuse and hard to localize (Gebhart \& Bielefeldt, 2016). No bedside clinical test package reliably identified the source in a diagnostic-accuracy systematic review against reference-standard blocks. Across 62 studies only three physical-examination findings were informative at all: the centralization phenomenon for the disc, and, for the sacroiliac joint, absence of midline pain plus a combination of provocation tests. No finding was informative for the facet joint (Hancock et al., 2007; Han et al., 2023). Imaging is the partial exception, and a narrow one. For the disc as the source, Modic type 1 vertebral endplate/marrow signal change on MRI carries a positive likelihood ratio of 10.00 (95\% CI 4.20–23.82) and Modic type 2 carries 8.03 (3.23–19.97). Each pairs near-total specificity with sensitivity near 12\%, so a negative scan excludes nothing (negative likelihood ratios 0.84 and 0.88; Han et al., 2023). These are rule-in tests, and they were estimated almost entirely in tertiary-care patients already referred for discography or injection. Degenerative findings are meanwhile common in people with no pain at all. Of 98 adults without back pain, 64\% had at least one intervertebral disk abnormality on MRI. Jensen et al. (1994) concluded from this that finding a bulge or protrusion in a patient with low-back pain “may frequently be coincidental”. A positive scan therefore localizes only against a compatible clinical picture. Low-back pain is consequently classified as non-specific in the great majority of cases, commonly cited as 90\%. In a primary-care series of 1,172 consecutive patients with acute low-back pain, fewer than 1\% had an identifiable specific cause (Maher et al., 2017). These two figures use different denominators and definitions and are not complements.

That number needs qualifying, and then quantifying. First the qualification: it holds for location \emph{alone}, and dynamic clinical assessment does better. Pain retreating toward the midline under repeated movement is the centralization phenomenon. It modestly raises the likelihood of discogenic pain — pooled positive likelihood ratios of 2.8 (95\% CI 1.4–5.3; Hancock et al., 2007) and 3.06 (95\% CI 1.44–6.50; Han et al., 2023), intervals wide enough that this is a contributing sign rather than a standalone test (Hancock et al., 2007; Han et al., 2023), and — a different virtue — is prognostically valid in 21 of 23 studies (May \& Aina, 2012). We state that number with its uncertainty, because the uncertainty is wide and tells against our own claim. Hancock's pooled estimate is LR+ 2.8 with a 95\% CI of 1.4–5.3. The lower bound therefore lies \emph{below} the threshold of 2 that the same review prespecified for calling a test informative at all. Centralization is the best single piece of evidence that provocation adds spatial information, and it is not a strong one. Provocation \emph{in combination} is firmer. Three or more positive sacroiliac provocation tests pool to LR+ 3.2 (95\% CI 2.3–4.4). That lower bound lies comfortably above the threshold, in the same review that found single manual tests uninformative in isolation (Hancock et al., 2007). That contrast, several weak rows becoming one informative test, is the empirical shape of the argument that follows.

Second, the companion paper formalizes location as a linear inverse problem (§4) and puts a number on the ambiguity. The \emph{identifiability floor} is the dimension of the referral matrix's near-null-space: the source combinations that produce nearly the same felt location. Provocation testing appends measurement rows, and informative rows \emph{lower} that floor. Lower, not abolish. The toy model recovers all five sources only because its provocation rows are built to be informative. The clinical rows deliver an LR+ of about 3, which shifts a diagnosis without settling it. The model explains why provocation helps. It does not license a claim that provocation identifies the source. One specific result would refute the multiplexing account. Suppose provocation testing added no diagnostic information over location alone, so that the combination performed no better than its parts. Location's failure here would then not be an identifiability problem, and another mechanism would explain it (Table 3).

\begin{mdframed}[backgroundcolor=boxbg,linecolor=boxln,linewidth=0.4pt,roundcorner=3pt]\textbf{Intuition.} Picture several water pipes that all leak into the same ceiling stain. The stain proves there is a leak but not \emph{which} pipe. No amount of staring at the stain will tell you, because several pipes produce the identical mark. You have to \emph{test the pipes one at a time} — send a surge of pressure through each in turn and watch which one makes the stain spread. The pipe that responds is the better candidate — though where several pipes respond to the same pressure change, the ambiguity stays. Location alone is the stain. Provocation testing is that surge: loading one candidate structure at a time, such as a straight-leg raise for the disc or compression for the facet joint, to see which reproduces the pain. (Psychology: convergent projection, poor spatial resolution. Medicine: non-specific pain, many generators in one region. Mathematics: for a continuous source, a rank-deficient inverse problem whose null space is exactly the set of indistinguishable causes.)\end{mdframed}

\begin{figure}[H]\centering
\includegraphics[width=\linewidth]{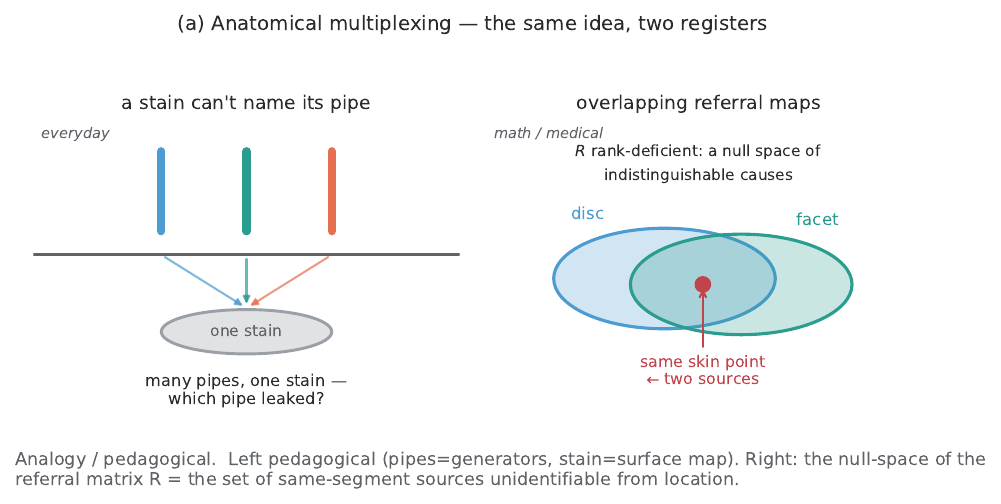}
\caption{The same idea in two registers. Everyday: several pipes leaking into one ceiling stain — the stain proves a leak but cannot name the pipe, and only pressurising each pipe in turn identifies it. Math/medical: overlapping referral maps, where one skin point is compatible with a disc and a facet source — the null space of the referral matrix $R$ is exactly the set of same-segment sources that location cannot separate. [Analogy on the left; the right panel states the null-space reading of $R$ exactly.]}
\end{figure}

\subsection{(b) Delocalized amplification}

Here the pain is not a corrupted localization but a change of generative model: it is produced centrally rather than by a peripheral lesion. “Central sensitization” is the clinical framework that describes presentations of this kind, and \emph{nociplastic pain} is a current mechanistic descriptor for pain not adequately explained by nociceptive or neuropathic mechanisms; it is not a synonym for the older literature's “functional” or “psychogenic” labels (Kosek et al., 2016). The entity is contested. A causal-criteria review found no study showing that it occurs in humans or that it causes chronic pain there, and judged questionnaire measurement circular (Velasco et al., 2024). The target here is instead an observable \emph{pain-behaviour pattern}: focal, stable, and provokable versus widespread, symmetric, and migrating, validated against outcomes, not against a quantitative sensory testing surrogate. The neural-field model below attaches to that pattern, the observable spatiotemporal behaviour of the pain map, and not to the contested construct. Central gain $s$ is a model parameter, not a claim that “central sensitization” has been measured.

The pattern has two signatures, one spatial and one temporal. Spatial extent covaries with questionnaire-measured central sensitization (Balasch-Bernat et al., 2022), though not with direct psychophysical measures in that cohort. Without sensitization, the same region shows no such pattern. Fifty-two patients with chronic unilateral nociceptive shoulder pain all scored below the study's own Central Sensitization Inventory cutoff of 35. In them, two-point discrimination — the smallest spacing at which two touches are felt as two, so a larger threshold means poorer tactile spatial acuity — did not differ between the painful and the pain-free shoulder, and extent was unrelated to tactile acuity (\emph{p} = .44; Caseiro et al., 2021). Caseiro et al. read those two nulls as evidence that body-schema disruption depends on which pain mechanism dominates. \emph{Temporal instability} is invisible to a single assessment: the pain map drifts from day to day. A neural-field model (companion §7, Figure 2) shows that this instability follows from a disinhibition-driven bifurcation rather than being assumed. A physicist would recognize the analogy to a phase transition, where a system changes character qualitatively once one parameter crosses a threshold. The analogy is structural rather than an identity, because a phase transition also carries thermodynamic-limit structure that a bifurcation need not. Below a critical gain, activity is a stable localized bump (focal pain); above it, spatial extent jumps to a widespread state. The two parameters play different roles and are easy to swap. Central gain is the \emph{control} parameter, the knob the system is turned by, as temperature is for boiling water. Spatial extent is the \emph{order parameter}: the quantity that changes character once the knob passes its critical value. Temporal instability is predicted to behave differently again. A decline in the high-gain regime follows from the model under five sufficient conditions. The interior maximum at the crossing is real in the approximation and negligible in the process: simulating the reduced dynamics directly bounds the climb into that maximum at 2.2\% against a decline past it of at least 29.5\%, so the falling limb, and not the peak, is what the model licenses this paper to claim (Table 3; companion §7.1). Cheap behavioural measures track static threshold sensitization better than they track dynamic sensitization. Pooled across 33 studies and 3,314 participants, the Central Sensitization Inventory correlates with pressure-pain threshold at $r$ = $-$.22. It correlates with temporal summation and conditioned pain modulation at only .10 and $-$.09 (Neblett et al., 2024). Two features of that meta-analysis matter more than its significance levels, which come cheaply at that sample size. No modality exceeds |$r$| = .27. In healthy controls the pressure-threshold association disappears entirely ($r$ = .002). The questionnaire is therefore a \emph{weak} proxy for psychophysical sensitization. \textbf{Pitfall:} central amplification should not be inferred from a questionnaire alone, both because that association is weak and because the construct itself is contested. Under the operational framing this is expected. The phone measures \emph{are} the pattern, not a proxy for a contested physiology.

The extent–sensitization association does not replicate. In a larger cohort of 146 previously hospitalised COVID-19 survivors with post-COVID pain, extent showed no significant association with the same questionnaire, nor with neuropathic-symptom, cognitive or psychological measures. Its association with pain \emph{intensity} ran in the opposite direction to the frozen-shoulder cohort ($r$ = $-$.201, \emph{p} = .014; Fernández-de-las-Peñas et al., 2022). Two cohorts, two different signs. We report it because it is the kind of finding a framework is tempted to leave out, and because it counts against this paper's model rather than around it. Extent is the \emph{order} parameter, and the model has it rise monotonically with central gain through the transition and keep rising past it (companion §7.1). A rank correlation detects a monotone rise. A reader could try to defend the null by appealing to curve shape, because a symmetric peak returns $r_s$ $\approx $ 0, but that defence is not available here: the peak the model discusses is in temporal instability, a different quantity, and not in extent. \textbf{Pitfall:} do not read this null as uninformative. A null between extent and sensitization is weak evidence \emph{against} the predicted monotone rise. The null is weak for reasons that have nothing to do with the shape of the curve — one cohort, a questionnaire-defined gain axis that tracks psychophysical sensitization only loosely, and a single pain drawing of a quantity the model says drifts between days — but weak evidence against a prediction is still evidence against it. What the pair of studies establishes more firmly is the sign reversal itself: a single linear summary of pain extent is not a stable quantity across populations, which is the structural claim this section makes. The design that would measure extent and gain well enough to settle the direction is preregistered in the supplement.

Population estimates inherit a measurement problem from that drift. If the defining feature here is \emph{movement}, extent and location drifting across days, then a study that assesses location once cannot see it. Prevalence estimates, treatment-response rates and the correlations above all rest on single assessments of a quantity the model says is non-stationary, so each averages over an unmeasured trajectory. That is tractable, not fatal, but it means the literature cannot yet settle how common this mechanism is. One specific result would refute the amplification account: if finite-window migration did not decline on the high-gain side of the transition, in a study with the power to see a decline of the predicted size, the bifurcation is the wrong structure. Extent rising monotonically with sensitization would \emph{not} refute it — that is what the model predicts extent to do — so the refutation rests on the migration limb alone. That conditional decline is the paper's least-supported claim and the one it most wants tested (Table 3).

\begin{mdframed}[backgroundcolor=boxbg,linecolor=boxln,linewidth=0.4pt,roundcorner=3pt]\textbf{Intuition.} Here the trouble is not a broken part sending a faithful signal from one place — it is the body's \emph{volume knob} turned up across the board. Turn it past a tipping point and the pain stops sitting still: it spreads and migrates. So the tell-tale is not \emph{where} the pain is on any single day but \emph{how it moves} across days. (Psychology: a delocalized, gain-amplified percept. Medicine: central sensitization / nociplastic pain. Mathematics: the system crosses a bifurcation — a qualitative change of regime, not a bigger version of the same thing.)\end{mdframed}

\begin{figure}[H]\centering
\includegraphics[width=\linewidth]{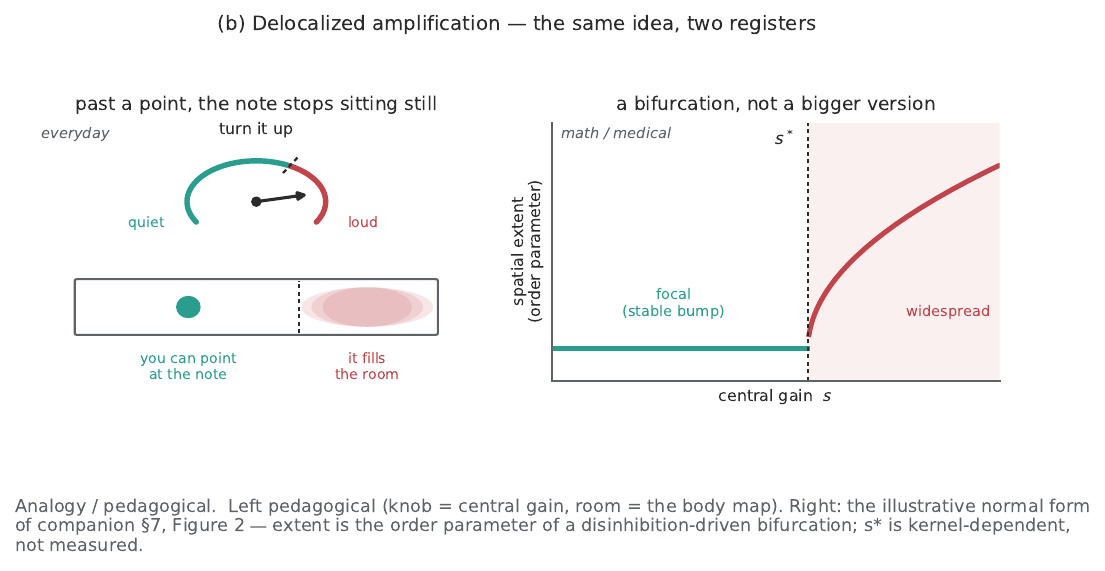}
\caption{The same idea in two registers. Everyday: a volume knob turned up until the note stops being placeable and fills the room. Math/medical: spatial extent as the order parameter of a disinhibition-driven bifurcation — focal below the critical gain $s^{*}$, widespread above it and rising further past it. [Analogy — the normal form is the illustrative one of companion §7; $s^{*}$ is kernel-dependent, not a measured value. Temporal instability is a different quantity from extent: the model licenses a decline on the high-gain side, not a detectable peak at the crossing (companion §7.1).]}
\end{figure}

\subsection{(c) Group-conditional presentation shift}

Location is hypothesized to be displaced from the source in a predictable way that depends on characteristics of the patient, the canonical illustration being the atypical presentation of myocardial infarction in women. Even the “typical” pattern discriminates weakly: oppressive chest pain carries a sensitivity of roughly 60\% and a specificity of roughly 58\% for acute infarction (Bruyninckx et al., 2008), so the diagnostic burden falls on objective testing.

\textbf{This mechanism's core claim — that felt or reported pain location itself shifts by patient group — has not yet been tested directly, and is marked \emph{open} rather than supported (Table 3).} A direct test needs group-stratified spatial or symptom-location data under a fixed reference diagnosis: do class-conditional location distributions actually differ by group, once the diagnosis and clinical setting are held fixed? No study cited below does that. What the literature below establishes instead, and establishes well, is a separate, downstream finding: an improvement in \emph{detection} does not by itself repair \emph{outcomes} when the \emph{treatment} step is the bottleneck. That pattern is real and clinically important, and the paper reports it in full, but it is evidence about a care pathway, not about where pain is felt or said to be. The companion paper keeps the two apart mathematically (§6): a group-conditional term in the location distribution $p(\ell  | s, g, z)$ is a different statistical object from a group-conditional term in the downstream decision rule $a^{*}(r, g, z)$ or in treatment uptake $p(t | a, s, g, z)$.

\textbf{A downstream corollary: detection is not the same as outcome.} \emph{Corollary} here names a companion result in the care pathway, not a logical consequence: a group-conditional shift in felt location would not by itself imply that detection-only intervention changes outcomes little, which depends besides on baseline detection, treatment uptake, treatment efficacy and the outcome model. The two are supported by different evidence and scored separately (Table 3). Sex-specific high-sensitivity troponin thresholds approximately doubled myocardial-infarction \emph{diagnosis} in women, from 11\% to 22\% (Shah et al., 2015); later evidence agrees that sex-specific and uniform thresholds perform comparably at presentation (Li et al., 2024). Neither study measures where the pain was felt or reported — both are about a biomarker threshold's diagnostic yield. Where that detection layer alone was fixed, clinicians still undertreated women — 15\% revascularized against 34\% of men — and one-year recurrent infarction or cardiovascular death barely moved, from 18\% to 17\% (adjusted hazard ratio 1.11, 95\% CI 0.92–1.33, where 1.0 would mean equal event rates, \emph{p} = .289; Lee et al., 2019). \textbf{Pitfall:} where the treatment step lags for a group, detection gains do not transfer to that group, so a detection fix reported as an outcome overstates what was achieved; and neither this nor the two studies below tests whether the underlying symptom \emph{location} itself differs by group — they test what happens once a case is identified.

A decision-theoretic model (companion §6) formalizes this as a standardized policy contrast: raising detection while holding treatment uptake at its currently observed, group-specific level changes outcomes little when that uptake is low. In the calibrated illustrative model the contrast is near zero for two reinforcing reasons: the treatment lever is weak (few of the newly detected are revascularized), and detection under the group-blind rule is already near-complete, so there is little detection left to recover. Both point the same way — the benefit depends on treatment, not on reclassification — but the second is a property of this calibration, not a general law. This is not automatically a \emph{controlled direct effect} in the causal-inference sense; that label requires potential outcomes and a fixed treatment level, assumptions the trials below do not establish (companion §6).

Detection gains \emph{can} translate into outcome gains where the treatment lever is not the bottleneck. The age-adjusted D-dimer for pulmonary embolism is that case. Among patients 75 or older with a non-high clinical probability, the age-adjusted cutoff raised the proportion in whom embolism could be excluded from 6.4\% to 29.7\% (95\% CI 26.4–33.3) — a more than fourfold gain — and added no false negatives (Righini et al., 2014). Lumbar imaging shows the same split as the troponin case, and one randomised trial measured both sides of it at once. Primary-care patients assigned rapid MRI rather than radiographs had far more anatomy detected: disk herniation in 33\%, disk bulge in 60\%, and central spinal stenosis in 20\%, none of which radiography can register at all. Twelve-month disability was nonetheless unchanged (modified Roland 9.34 vs 8.75; mean difference $-$0.59, 95\% CI $-$1.69 to 0.87; Jarvik et al., 2003). The outcome null is not peculiar to that trial. Six further trials of immediate lumbar imaging against usual care without it find no difference in pain or function, at either short- or long-term follow-up (Chou et al., 2009). Both MRI and imaging trials are further evidence \emph{for the corollary} — better detection without an outcome lever — and neither one, on its own or together, tests group-dependent location displacement; the low-back-pain trials do not even stratify by group. They are retained here as a second setting for the corollary, not as a second setting for the location hypothesis, and Table 3 scores them accordingly.

Because the bias is group-conditional and not random, a pooled accuracy figure is the wrong summary: it averages over the very groups that differ, and can conceal poor performance in one or more groups, and materially different error profiles across them. The matching instrument for the \emph{detection} layer is a group-stratified accuracy analysis: likelihood ratios reported \emph{by} the covariate, not collapsed across it, paired with a policy analysis separating detection from treatment uptake, since only the second carries the outcome benefit. \textbf{What would support the location-displacement hypothesis itself} is direct: group-stratified pain-location or symptom-site data, collected under a fixed reference diagnosis, showing that class-conditional location distributions differ by group after reporting and measurement differences are modeled. Absent that, the honest status is open. \textbf{What would refute the corollary} is equally direct: if repairing detection alone improved outcomes, the policy contrast would not be near zero. Two settings currently say otherwise (Table 3).

\begin{mdframed}[backgroundcolor=boxbg,linecolor=boxln,linewidth=0.4pt,roundcorner=3pt]\textbf{Intuition.} The corollary, not the hypothesis, is the load-bearing part here so far. A better detector finds more true events. But if the \emph{treatment} step still lags for a group, finding the event sooner changes little for that group — two locks on one door; opening one leaves the door shut. Whether the pain itself is felt or reported in a systematically different place for that group is a separate, still-open question this account does not yet have direct evidence for. (Psychology: population/observer differences in the report — hypothesized, not yet shown. Medicine: atypical presentation and a demographic detection/treatment gap — shown. Mathematics: a group-conditional term in the location distribution is open; the policy contrast at the decision threshold, downstream of it, is not.)\end{mdframed}

\begin{figure}[H]\centering
\includegraphics[width=\linewidth]{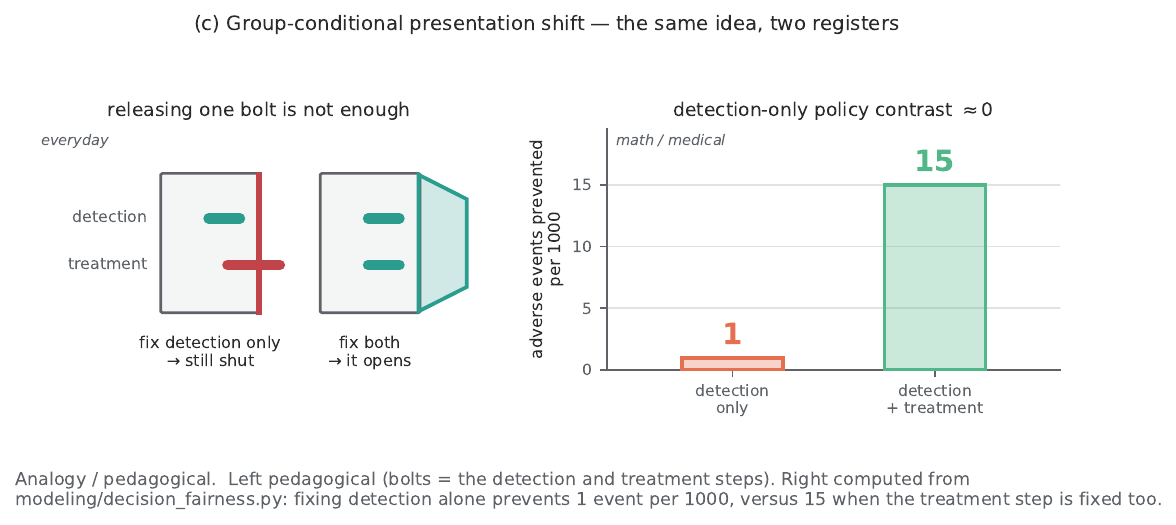}
\caption{The same idea in two registers. Everyday: one door with two bolts — releasing the detection bolt alone leaves it shut; release both and it opens. Math/medical: the standardized detection-only policy contrast, as events prevented per 1000 — 1 from fixing detection alone, 15 when the treatment step is fixed too. [Analogy on the left; the right panel is computed from the mediation model of companion §6.]}
\end{figure}

\section{Synthesis: One Inference, Distinct Nodes}

The distinction is not only clinical but mathematical. Formal models place each mechanism in a different branch of the field: multiplexing is a linear inverse problem, amplification is generative-model selection resolved by a temporal information channel, and displacement is a group-conditional term in the perceptual model, kept separate from the decision and treatment layer that follows it (companion §3). One Bayesian \emph{generative} model unifies the first three: recovering the cause from the report. It fails at a different node in each case: the likelihood in multiplexing, the generative-model class in amplification, the group-conditional percept distribution in displacement. Decision and treatment are not additional nodes in that probability model; a loss function is not a probability factor, so clinical action, treatment uptake, and outcome are represented as a separate layer that reads the posterior out (companion §3). Keeping the probability model and the decision layer apart stops a downstream treatment disparity from being read as a perceptual displacement, the error that mechanism (c) above corrects for. A fourth failure occurs at the observation model, where expectation or the measuring instrument distorts the report; the companion paper formalizes it as the spatial Bayesian mislocalization model. The phenomenon is not hypothetical, though the particular decomposition remains untested. Chronic low-back-pain patients show measurable mislocalization that is dissociated from pain intensity (Wand et al., 2013). The neural substrate is partly differentiated as well: the posterior insula supports sensory-discriminative spatial coding, the anterior insula affective salience (Lu et al., 2016). That is consistent with partial separability of the spatial and affective components; it does not by itself establish that a clinical localization error is independent of distress. This generative model subsumes the separate perceptual results: mutual information is a property of the graph, and the inverse problem is its likelihood node. A single principle follows for a \emph{fixed} target: retaining another observation cannot reduce recoverable information about it — an exact identity (the mutual-information chain rule), not an empirical claim that repeated assessment always gives a better estimate than a single one (Figure 3; companion §5).

\begin{figure}[H]\centering
\includegraphics[width=\linewidth]{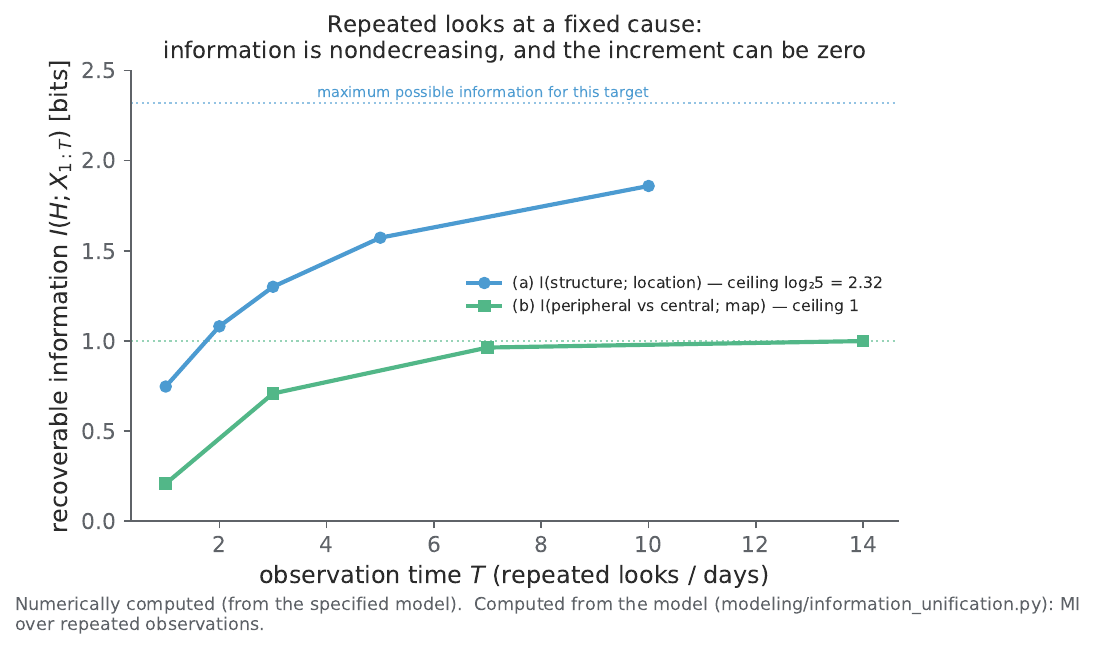}
\caption{Figure 3. Additional observations cannot reduce information about a fixed target. Recoverable information about a fixed cause is nondecreasing with observation time — an exact identity (the mutual-information chain rule, companion §5), not an empirical claim that longitudinal tracking always beats a snapshot in practice. What to look at: the curve never turns down and can plateau, so cumulative information after a second look is never lower than after the first — though the second look's own increment can be small or zero, and may not be worth its collection cost. [Numerically computed — Monte-Carlo mutual information over the discrete repeated-observation channel of companion §5 (categorical site draws for (a), Bernoulli day-maps for (b)); not a Gaussian channel.]}
\end{figure}

The graph does not by itself separate delocalized amplification from the reporting node. A delocalized, gain-amplified percept and a report pulled onto a diffuse cognitive prior can produce the \emph{same} static pain map: a widespread, hard-to-localize drawing. The nodes are distinct in the model but observationally degenerate on a single snapshot. Separating them is therefore not a matter of reading one map more carefully. It requires the dynamic and provocation tests below: spatial \emph{instability} over time for amplification, and a zero-expectation control trial for the reporting node.

\begin{mdframed}[backgroundcolor=boxbg,linecolor=boxln,linewidth=0.4pt,roundcorner=3pt]\textbf{Intuition.} Think of a single machine that turns \emph{what is actually wrong} into \emph{where it hurts}. Each of the three failures breaks a different part of that machine. Multiplexing blurs the lens, so the machine cannot resolve which source the signal came from. Amplification swaps in the \emph{wrong internal model} of what is generating the pain. Displacement mis-sets the readout dial, and by a different amount in different patient groups — and this third one is what we propose, not what anyone has yet shown. The literature shows something narrower, one step downstream: fixing detection often leaves treatment uptake untouched.\end{mdframed}

\textbf{Pitfall:} one pooled “diagnostic utility” number blurs all three failures together, so it measures the symptom \emph{in that population under that mechanism} rather than the symptom itself.

A snapshot cannot reveal how the symptom behaves over time. Each fresh look at the pain can only add information about a fixed cause, never remove it, though the increment can be tiny; whether repeated assessment is worth collecting depends on that increment against its cost, not on the sign of the theorem alone. (Psychology: the temporal structure of perception. Medicine: longitudinal symptom tracking. Mathematics: repeated use of a communication channel monotonically raises the recoverable information, but does not bound the size of any one increment.)

\begin{figure}[H]\centering
\includegraphics[width=\linewidth]{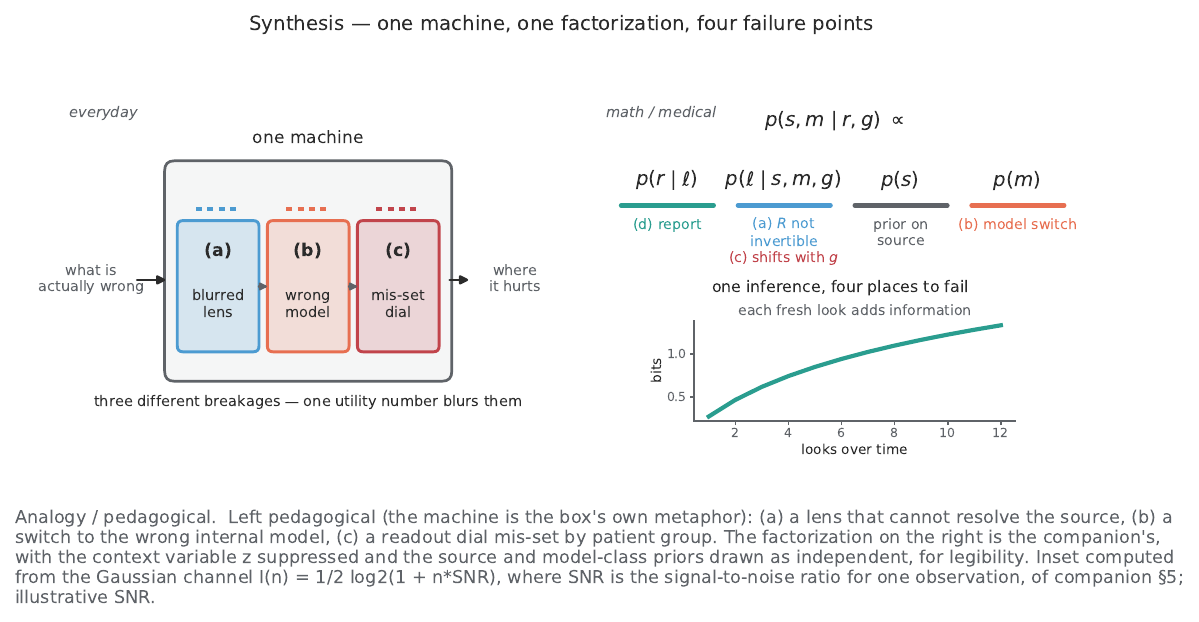}
\caption{The same idea in two registers. Everyday: one machine turning what is actually wrong into where it hurts, with three parts that can break — a lens that cannot resolve the source (a), a switch to the wrong internal model (b), and a readout dial mis-set by patient group (c). Math/medical: the same three, plus the report node (d), flagged on the factors of a single Bayesian factorization — the paper's claim that these are not three theories but one inference failing in four places. [Analogy on the left; the factorization is exact, and the inset is computed from the Gaussian channel of companion §5.]}
\end{figure}

\section{Model Predictions and Current Evidence}

Each mechanism makes a distinct, testable prediction, and the models make those predictions concrete across a range of conditions (Figure 4). For anatomical multiplexing, informative provocation rows \emph{lower} the \emph{identifiability floor}, the number of source combinations location alone cannot tell apart. For delocalized amplification, finite-window spatial migration \emph{declines} once central gain carries the system onto the high-barrier side of the transition: simulating the reduced dynamics bounds the climb toward the interior maximum at 2.2\% against a decline past it of at least 29.5\%, so the falling limb rather than a peak is the empirical target (companion §7.1). The testable form of that is a pair rather than a single measure: extent rises straight through the transition and keeps rising, so extent alone cannot tell the crossing from the saturated state, while extent together with \emph{how it changed between two visits} can (companion §7, with the control in that section on what the second variable does and does not measure). For displacement, the location hypothesis predicts group-conditional differences in felt or reported location, which is not yet directly tested; its downstream corollary is that fixing detection alone barely moves patient outcomes, because the benefit is carried by the treatment step downstream of detection. For the reporting node, an expectation that sharpens, which is a reporting prior of rising precision, pulls the reported location toward itself and away from where the pain is felt. The evidence for and against these predictions is uneven, and Table 3 states the balance honestly; the supplement carries the maintained version, updated as evidence arrives.

\begin{figure}[H]\centering
\includegraphics[width=\linewidth]{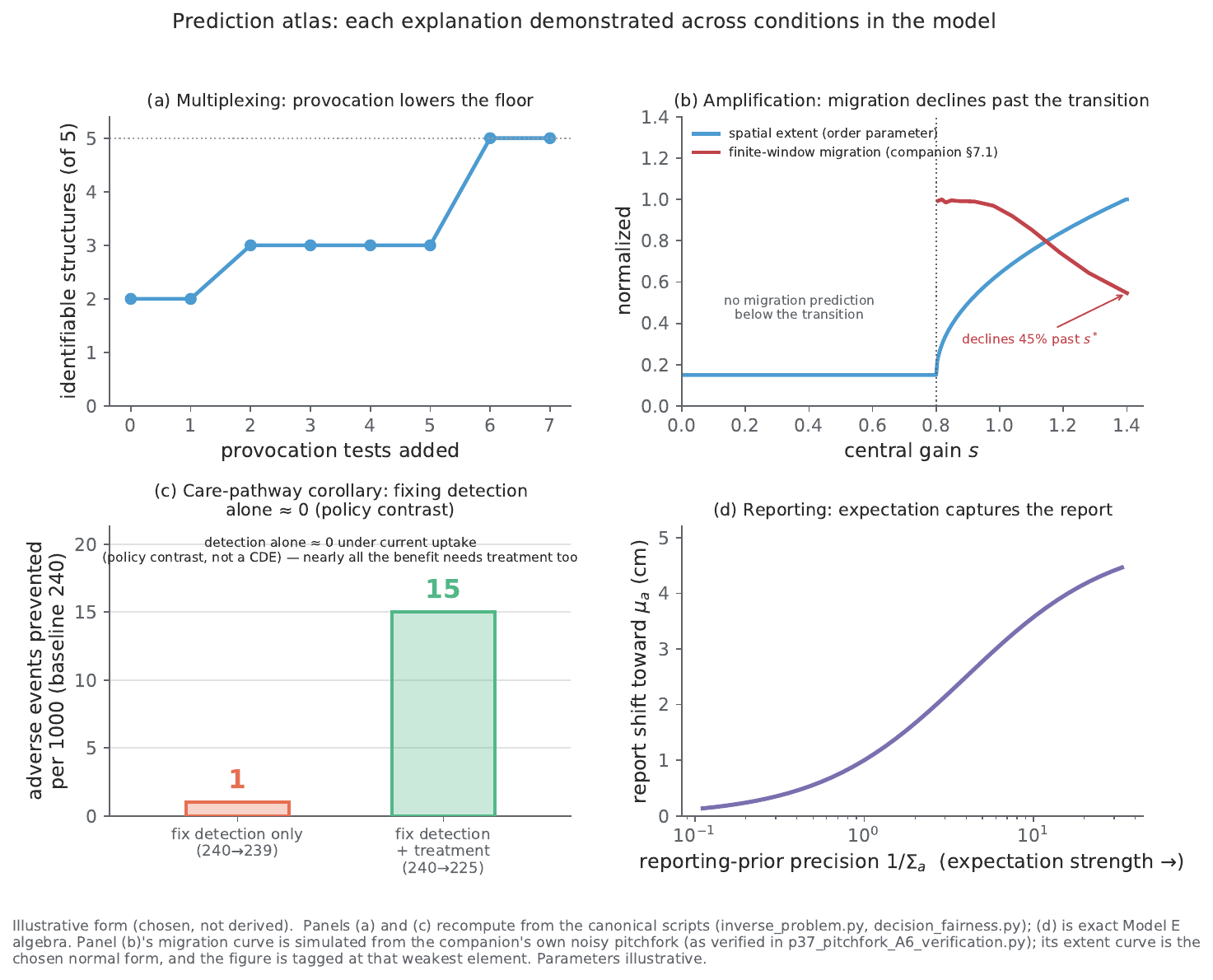}
\caption{Figure 4. What each failure predicts, once it is written as a model. Four panels: the three perceptual failure points, and for displacement its downstream care-pathway result rather than its own location prediction, which is open and has no data to plot. Adding provocation tests to a many-to-one map can reduce the identifiability floor that finer localization cannot (a). Finite-window migration is predicted to \emph{decline} once central gain carries the system past the transition, even while extent keeps climbing (b) — the falling limb is the falsifiable feature, and an intermediate peak is not predicted (companion §7.1). Improving detection while leaving the treatment step untouched leaves outcomes close to where they started (c) — the care-pathway companion result, not the group-conditional location shift itself. An expectation, once confident enough, drags the reported location onto itself (d). The point of the set is that the four predicted shapes differ from one another, so a study can tell them apart. [Numerically computed — each panel computed from its model; parameters illustrative.]}
\end{figure}

Which mechanism dominates also predicts how well location performs in a familiar case. Diagnostic value runs from strongly localizing — right-lower-quadrant \emph{pain} in appendicitis, positive likelihood ratio 7.31–8.46 (Wagner et al., 1996, as tabulated by Yeh, 2008) — to actively misleading, as in atypical myocardial infarction. That figure is a range across heterogeneous studies rather than a confidence interval, because the source could not pool them. It also belongs to the patient's \emph{report} of where it hurts, not to a palpation sign: the examination signs in the same review are far weaker (rebound tenderness 1.10–6.30, rigidity 3.76). Delocalized amplification (b) is uniformly poorly localizing, multiplexing spans the range, and displacement splits by presentation.

The status column uses a fixed vocabulary, in descending order of evidential strength:

\begin{itemize}\setlength{\itemsep}{2pt}\item \textbf{\emph{supported}}: at least one controlled or prospective study matches the prediction's direction.\item \textbf{\emph{consistent}}: observational or diagnostic-accuracy data align, but no study was designed to test the prediction.\item \textbf{\emph{open}}: a decisive study is designed but not yet run.\item \textbf{\emph{untested}}: no data of the needed form exists.\end{itemize}

\emph{Table 3. Predictions and their current evidential status.}

{\footnotesize\begin{xltabular}{\textwidth}{>{\raggedright\arraybackslash}X>{\raggedright\arraybackslash}X>{\raggedright\arraybackslash}X}
\toprule
Mechanism / model & Key prediction & Status \\
\midrule
\endhead
Anatomical multiplexing (a): inverse problem & location alone cannot identify the source; provocation adds source-discriminating information & consistent \\
Delocalized amplification (b): bifurcation & finite-window migration \emph{declines} on the high-gain side of the transition; an intermediate peak is \textbf{not} predicted & open — key test designed; derived under five sufficient conditions (companion §7.1), the open premise being that the destabilizing mode is a displacement mode \\
\textbf{Displacement (c): location hypothesis} & felt/reported location itself differs by patient group, holding diagnosis and setting fixed & \textbf{open — no direct test yet (revised from “supported”)} \\
Care-pathway corollary (downstream of c) & fixing detection alone gives $\approx $ 0 outcome benefit when treatment uptake is the bottleneck & supported (two settings) \\
Reporting node (d) / Model E & for an unconstrained continuous source, identifiable only if the whitened \emph{R} — the referral matrix rescaled by measurement noise — has &  \\
Additional observation (chain rule) & retaining another observation cannot reduce recoverable information about a \emph{fixed} cause & exact identity — proved, not an empirical prediction; practical value of serial tracking is untested \\
Whole framework & the three are separable by \emph{instrument}, not by appearance & untested — falsifier stated \\
\bottomrule
\end{xltabular}}

The evidence behind each row, in the same order.

For \textbf{multiplexing}, bedside tests cannot identify the disc, facet-joint, or sacroiliac source (Hancock et al., 2007; Han et al., 2023). Provocation helps without solving the problem. Centralization carries a positive likelihood ratio of 2.8 (95\% CI 1.4–5.3), and the lower bound of that interval falls below the review's own informativeness threshold. Three or more positive sacroiliac provocation tests pool to 3.2 (2.3–4.4). The status is \emph{consistent} rather than supported for two reasons: the strongest supporting likelihood ratio is modest, and the referral matrix \emph{R} has never been estimated from data.

For \textbf{amplification}, tactile acuity is reduced in chronic low-back pain. Two independent meta-analyses converge on the size of that reduction: 11.7 mm, 95\% CI 5.5–17.8 (Catley et al., 2014), and 11.74 mm, 95\% CI 8.61–14.87 for lumbar-specific measurement (Adamczyk, Luedtke, \& Saulicz, 2018).

The pooled estimates are weaker than they look, and their caveats matter more than the numbers. Both reviews report the same measurement failure, independently. Catley et al. find the assessor unblinded in 94\% of the pooled studies and no study reporting reliability indices, and judge their own estimate as possibly overstating the disparity and possibly publication-biased. Adamczyk, Luedtke, and Saulicz find reliability and blinding below an acceptable score in 84\% of their included studies, with a single case-control study blinding its assessor. They ask that the outcome be read with caution until better designs replicate it. Excluding the higher-risk studies shrinks Adamczyk's effect to 9.49 mm (95\% CI 3.64–15.34). That smaller estimate is the less stable one: heterogeneity rises from I$^{2}$ = 34\% to I$^{2}$ = 80\%, conventionally read as very high heterogeneity, and the interval nearly doubles in width (Adamczyk, Luedtke, \& Saulicz, 2018). A third review looked at the instruments rather than at the effect, and landed in the same place. Across ten studies of six sensorimotor tests in chronic low-back pain, two-point discrimination separates patients from healthy controls on moderate evidence. But the evidence bearing on its \emph{reliability} was too poor in quality to assess at all, so the authors ask that the validity finding be read cautiously (Ehrenbrusthoff et al., 2018).

\textbf{Reduced acuity is also not specific to the chronic, centrally amplified case, and that matters more than the pooled size.} In a double-blind experiment, saline-induced acute \emph{nociceptive} low back pain in healthy volunteers made two-point discrimination worse within the session, from 56.94 mm to 64.22 mm, with maximum reported pain predicting the change ($\beta $ = 0.55, 26\% of the variance; Adamczyk, Saulicz, et al., 2018). Ordinary nociception degrades the very measure the chronic literature reads as a central signature. A raised threshold therefore does not by itself say which mechanism produced it — which is this paper's thesis about pooled numbers, arriving here from the measurement side. A distress confound remains unruled-out (Christensen et al., 2017). And the migration-decline claim itself — the actual prediction — is untested.

For the \textbf{displacement location hypothesis}, no cited study measures group-conditional felt or reported location under a fixed reference diagnosis, so the status is open rather than supported; this is a change from the manuscript's earlier draft, made because the studies previously entered here test diagnostic reclassification and treatment uptake, not location. For the \textbf{care-pathway corollary}, two settings show detection improving while outcomes did not. In the high-sensitivity troponin rollout, the adjusted hazard ratio for the outcome was 1.11, 95\% CI 0.92–1.33, \emph{p} = .289 (Lee et al., 2019), which is no detectable benefit. In lumbar imaging, rapid MRI detected substantially more anatomy than radiography, and twelve-month disability was unchanged (Jarvik et al., 2003); the same outcome null holds across six further trials (Chou et al., 2009). That imaging row is the only one with no countervailing evidence entered against it — for the corollary specifically, not for the location hypothesis above it.

For the \textbf{reporting node}, mislocalization dissociates from pain intensity (Wand et al., 2013) and whole-brain pattern analysis decodes perceived location (Jung et al., 2019); both are qualitative support, the model is not yet fitted, and the identifiability risk needs a provocation pilot. The chain-rule identity for \textbf{additional observation} needs no empirical test — it is proved, not predicted — but its \emph{practical} value has no evidence either way, because no dataset yet tracks felt \emph{location} within patients at a daily timescale: longitudinal cohorts count pain \emph{sites} at survey waves instead of mapping location day to day (Kamaleri et al., 2009; Christensen et al., 2017), and the nearest momentary-assessment study tracks intensity, not location (Pagé et al., 2022). The \textbf{whole framework} is tested by a design, not by data: a large digital pain-drawing corpus, cross-referenced with the three instruments. That design carries its own falsifier. If a single continuous axis of spatial precision predicted every case, and no instrument partitioned them, the taxonomy would collapse back into the one-gradient account it argues against.

Read in that vocabulary, the summary runs as follows. The displacement location hypothesis is \emph{open}; its downstream care-pathway corollary is \emph{supported}, in two settings including a randomized trial — the two must be scored separately, which is the change from the manuscript's earlier draft. Multiplexing is \emph{consistent} with diagnostic-accuracy data. Amplification is \emph{open}: its decisive test is designed but not yet run. The novel spatial model of the reporting node is \emph{untested}, no data of the needed form existing yet, and the practical value of serial observation is likewise untested even though the underlying information identity is proved. The distress-confound test (Limitations) is, for this framework, the most decisive near-term study. None of the models has been \emph{fitted} to primary data; their predictions are nonetheless sharp enough to falsify. The three-way split is itself falsifiable: if no instrument partitions the mechanisms, and a single axis of spatial precision explains every case, the distinction reduces to the one gradient it set out to replace.

Table 4 takes the complementary, data-first view, asking which \emph{specific published measurements} either set a model parameter (calibration) or support a prediction, and which parameters have no published value yet. That list is what a fitting study would work from. Several components have published benchmarks that can inform priors or plausible ranges; the full parameter set remains unfitted. They are not free-floating.

\emph{Table 4. Published measurements that constrain, motivate, or test model components.}

{\footnotesize\begin{xltabular}{\textwidth}{>{\raggedright\arraybackslash}X>{\raggedright\arraybackslash}X>{\raggedright\arraybackslash}X>{\raggedright\arraybackslash}X}
\toprule
Published source & Reported value & Role & Parameter set / prediction supported \\
\midrule
\endhead
Schlereth et al., 2001 & cutaneous point-localization $\approx $ 5–9 mm & calibrates & $\Sigma _c$ (transmission noise); scopes the ambiguity to \emph{deep/visceral}, not cutaneous \\
Kurosawa et al., 2015 (primary abstract, verified) & referral zone \emph{location} shifts with the section stimulated: in 50 patients with block-confirmed sacroiliac dysfunction, the four posterior sections of the joint referred to four different buttock zones — “Dysfunctional upper sections of the SIJ are associated with pain in the upper buttock and lower sections with pain in the lower buttock” & calibrates & structure of the referral matrix $R$ — deep sources carry \emph{partial} spatial information \\
O'Neill et al., 2002 (primary abstract, verified); Graven-Nielsen et al., 1997 (primary abstract, verified) & referred \emph{extent} grows proximal$\to $distal with stimulus intensity: on intradiscal stimulation in 25 patients, “pain originated proximally and progressed distally as stimulus intensity increased”. Graven-Nielsen adds an intensity–extent \emph{correlation} in muscle (pain rating against referred area, \emph{R} = .47, \emph{p} < .0001, 231 observations) — \textbf{but its own experimental increase in referred area across sequential infusions was non-significant, and it never varies the level stimulated, so it supports the intensity clause only and by correlation only} & calibrates & same row: $R$ carries partial spatial information, and how far it carries scales with intensity \\
Han et al., 2023; Hancock et al., 2007 & centralization LR+ $\approx $ 2.8–3.1; Hancock's own pooled value is 2.8 (95\% CI \textbf{1.4}–5.3) & calibrates & informativeness of the provocation rows $P$ appended to $R$ (raises effective rank) — \textbf{and bounds it: the CI's lower limit is below the informativeness threshold of 2, so a single provocation row is less informative than the toy model's constructed rows} \\
Catley et al., 2014 & CLBP (4 studies, 5 comparisons): standardized effect 1.14 (95\% CI .54–1.74), i.e. \textbf{11.7 mm (95\% CI 5.5–17.8)}, a 26\% difference (95\% CI 12–39\%); I$^{2}$ = 69\% & calibrates & widening of $\Sigma _c$ under sensitization — but the authors state the estimate may \textbf{overestimate} the true disparity (94\% nonblinded assessors; no reliability indices reported; possible publication bias) \\
Adamczyk, Luedtke, \& Saulicz, 2018 (primary, read) & lumbar-specific replication (n = 547 / 346): 11.74 mm (95\% CI 8.61–14.87), I$^{2}$ = 34\%; \textbf{9.49 mm (95\% CI 3.64–15.34), I$^{2}$ = 80\%, excluding higher-risk-of-bias studies}; four case-control studies carry 73\% of the effect & calibrates & the two reviews converge on $\approx $ 11.7 mm independently. \textbf{The bias-sensitivity is not the reassuring bound it looks like:} the conservative estimate is also the heterogeneous one, and the authors ask that the outcome be read with caution until better-designed studies replicate it \\
Adamczyk, Saulicz, et al., 2018 (abstract verified) & healthy volunteers, saline-induced acute \emph{nociceptive} low back pain, double-blind: two-point discrimination \textbf{worsened} during pain, 56.94 mm $\to $ 64.22 mm; maximum pain predicted it ($\beta $ = 0.55, 26\% of variance) & \textbf{scopes} & amplification: acuity loss is \textbf{not specific} to chronic or centrally amplified pain, so a raised threshold alone cannot identify the mechanism. Strengthens the paper's own case against one pooled “utility” number, from the measurement side \\
viscerosomatic convergence (dual-input dorsal-horn neurons; animal) & a substantial fraction of dorsal-horn neurons receive both visceral and somatic input & calibrates & degree of many-to-one convergence in $R$, \textbf{animal only; no verified human figure, a gap} \\
Hancock et al., 2007 (primary, read) & with +LR > 2 / $-$LR < 0.5 as the prespecified informativeness threshold, \textbf{“none of the tests for facet joint pain were found to be informative”}; individual symptom- and location-based items are at +LR $\approx $ 0.6–1.5, e.g. absent centralization 1.1 (95\% CI 0.9–1.5) and 1.1 (1.0–1.4), traumatic onset 1.0–1.01; single manual SIJ tests likewise uninformative, whereas $\ge $ 3 positive SIJ provocation tests pool to +LR 3.2 (2.3–4.4), $-$LR 0.29 (0.19–0.44) & supports & multiplexing/(E) identifiability floor: same-segment sources unresolvable from location, and provocation rows $P$ lower the floor. \textbf{Note:} the summary figure “location-alone LR $\approx $ 1.0–1.1” is our \emph{derived} reading of these rows, not a pooled estimate Hancock reports \\
Fernández-de-las-Peñas et al., 2022 & post-COVID pain (n = 146): pain extent \textbf{not} associated with the Central Sensitization Inventory, and extent–intensity association \emph{negative} ($r$ = $-$.201, \emph{p} = .014), opposite in sign to Balasch-Bernat & \textbf{challenges} & amplification: the linear extent$\leftrightarrow $sensitization association does not replicate. Extent is the model's order parameter and is predicted to rise monotonically with gain (companion §7.1), a shape a rank correlation can detect, so this null is weak evidence \emph{against} the prediction rather than uninformative about it. Weak because of one cohort, a questionnaire gain axis and a single drawing — not because of curve shape \\
Caseiro et al., 2021 (primary, read) & chronic unilateral \emph{nociceptive} shoulder pain (n = 52), every participant below the study's own CSI cutoff of 35 (observed max 34; CSI Part B conditions excluded by design): two-point discrimination \textbf{no different} between the painful and pain-free shoulder (anterosuperior mean difference 0.5 mm, \emph{p} = .581; lateral 1.14 mm, \emph{p} = .173), laterality judgment no different, and pain extent unrelated to either (\emph{F} = 0.98, \emph{p} = .44) & supports & amplification by its \textbf{absence}: below the study's own sensitization cutoff, no acuity degradation. A null under that operational definition, not evidence that central gain is absent. The negative control for Balasch-Bernat in the same body region. Weak on its own — a null in a nociceptive sample is unsurprising — and load-bearing only as the contrast \\
Wand et al., 2013 & CLBP mislocalization dissociated from pain intensity & supports & node \emph{d}: a spatial disturbance distinct from intensity \\
Jung et al., 2019 & whole-brain multivoxel pattern analysis (MVPA) decodes perceived pain location — \textbf{but the task is narrow}: 14 healthy men discriminating two adjacent forearm needle sites, whole-brain accuracy 58.6\% against a 50\% chance level (univariate analysis failed), so this is acute experimental stimulation, not spontaneous or clinical pain & supports & (E) premise: a decodable spatial code exists to be biased \\
Lee et al., 2019 (High-STEACS) & detection recovered, outcomes unchanged (HR 1.11, 95\% CI 0.92–1.33, \emph{p} = .289) & supports & care-pathway corollary of (c): standardized policy contrast $\approx $ 0 under current uptake. Does \textbf{not} test location displacement itself \\
Jarvik et al., 2003 (primary, read) & rapid MRI vs radiographs, n = 380: detection up — disk herniation 33\%, bulge 60\%, central stenosis 20\%, all invisible to radiography — with 12-month disability unchanged (mean difference $-$0.59, 95\% CI $-$1.69 to 0.87) & supports & care-pathway corollary of (c): standardized policy contrast $\approx $ 0 — second setting, and the only source carrying \textbf{both} clauses inside one randomised trial. Does \textbf{not} test location displacement \\
Chou et al., 2009 (primary, read) & six trials, n = 1804: immediate lumbar imaging vs usual care, pain and function unchanged short- and long-term & supports & care-pathway corollary of (c): the outcome null generalises past a single trial. Speaks to outcomes \textbf{only} — the review excluded head-to-head modality comparisons by design, so it carries no detection claim (an earlier draft attributed one to it), and none of the six trials stratifies by group, so it does not test location displacement either \\
Goldreich, 2007 & Bayesian observer model of \emph{tactile} spatial perception & precedent & method precedent for Model E (touch, not pain — the novelty anchor) \\
\emph{— no published value —} & deep/visceral point-localization error (mm/cm) & \textbf{gap} & would set $\Sigma _c$ for \emph{deep} sources (the model's central premise) \\
\emph{— no published value —} & distress-adjusted extent $\to $ outcome & \textbf{gap} & separates amplification as signal (S1) from distress marker (S2) — the decisive test \\
\emph{— no published value —} & migration vs \emph{graded} sensitization & \textbf{gap} & tests the amplification migration-decline prediction (companion §7.1) \\
\bottomrule
\end{xltabular}}

\section{The Reporting Law Across the Senses: A Consilience Check}

The reporting node is not an arbitrary equation: $E[r] = \Sigma _k W_k x_k$ with precision weights $W_k \propto  \Sigma _k^{-1}$ is exactly the \emph{optimal cue-integration} rule, the maximum-likelihood way to combine noisy estimates: in plain language, the reported location is a weighted average of the available location cues, and the more reliable a cue is, the more weight it carries. That licenses a \emph{consilience} check, meaning a test of whether independent lines of evidence, gathered for unrelated reasons, converge on the same conclusion. Its limits belong up front. Precision-weighted integration is the best-established rule for how the brain fuses \emph{redundant} cues in vision, audition, and touch, and the review that consolidates it presents integration as one of several combination strategies, selected according to the type of information (Ernst \& Bülthoff, 2004). The pain-reporting node therefore inherits a well-tested \emph{form} rather than inventing one. It does not follow that pain reporting uses the same weights: the pain-specific $W$, $R$, and $\mu _a$ are unfit. Nociception may well depart from the tidy external-sense case, because it carries dense affective and visceral projections and because its anatomical multiplexing runs deeper. The other senses establish that this functional form is a well-tested benchmark for cue integration; whether pain reporting uses the same weighting rule is the empirical step still open (Figure 5). The claim is therefore shared structure, not out-of-sample confirmation in pain.

\begin{figure}[H]\centering
\includegraphics[width=\linewidth]{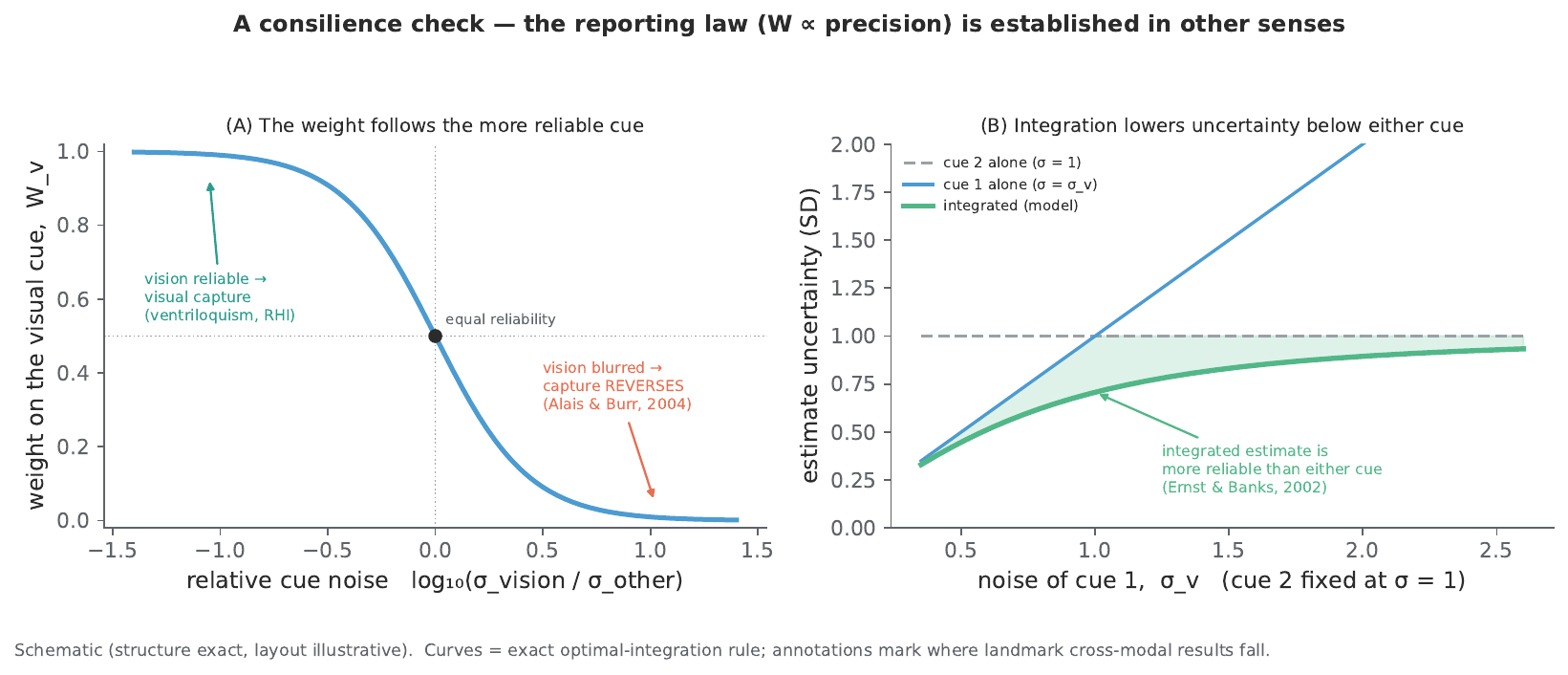}
\caption{Figure 5. A consilience check — the reporting law W $\propto $ precision is the optimal cue-integration rule, a \emph{form} independently established in senses the model was not built for (it does not by itself confirm the pain-specific weights). (A) The weight on a cue follows its relative reliability: when the usually-dominant visual cue is degraded past the other's reliability, capture reverses (the reversal of audiovisual ventriloquism; Alais \& Burr, 2004). (B) Combining two cues yields an estimate more reliable than either alone — the variance-reduction signature of statistically optimal visual–haptic integration (Ernst \& Banks, 2002). [Schematic — curves are the exact optimal-integration rule; annotations mark where the landmark cross-modal results fall. Data from Alais and Burr (2004) and Ernst and Banks (2002); the figure is original and reproduces no published figure.]}
\end{figure}

Three consequences follow, the first two directly from the weights and the third only once the causal-inference stage is added, and each is established in an adjacent sensory literature. First, the report follows the more reliable cue ($W_k \propto  \Sigma _k^{-1}$). Degrading the usually-dominant cue should therefore \emph{reverse} which cue captures — and it does. Blur a visual target until it is less reliable than the sound, and audiovisual ventriloquism flips from visual to auditory capture (Alais \& Burr, 2004). Visual–haptic size perception likewise weights each sense by its reliability, close to the statistical optimum (Ernst \& Banks, 2002; Figure 5A). Second, integrating two cues is more reliable than either alone (the precisions add) — the variance-reduction signature established in that same visual–haptic paradigm (Ernst \& Banks, 2002; Figure 5B). Third, capture releases when the cues are too discrepant to share a common cause, which is the model's \emph{common-cause} (Bayesian causal-inference) stage (companion §8, Figure 6). The rubber-hand illusion weakens as the fake hand moves farther from the real one (Lloyd, 2007), and is abolished under an anatomically implausible posture (Tsakiris \& Haggard, 2005). That same manipulation also \emph{reverses} the drift's sign instead of merely abolishing it, which release alone cannot explain. That is \emph{repulsion}, the report pushed \emph{away} from a cue rather than toward it, and it belongs with the boundary noted below. These last two are \emph{somatosensory} — touch on one's own body — and so are closest to pain. The audiovisual and visual–haptic cases, by contrast, match the mathematical \emph{form} but not the modality, and are the weaker, structural end of the argument.

Two caveats bound what this borrowing establishes: the results were not predicted in advance, and the linear blend cannot produce repulsion. First, none of these is a new experiment the model called in advance; they are established results whose \emph{form} the reporting law shares. That is consilience, not confirmation of the pain-specific weights, and least of all a licence to read visceral pain off an audiovisual study. Nor is the fit exact even in those adjacent literatures. Alais and Burr's own bimodal thresholds improve by a factor of 1.18–1.35 against a predicted ceiling of $\surd $2 $\approx $ 1.41, and one observer deviated significantly from optimality (\emph{p} = .04). The law is a close approximation outside pain rather than a perfect one, so a pain-specific version should not be expected to fit more closely. Second, the linear blend has a bound worth stating precisely, and it is narrower than it looks. Where the reporting weight $W$ acts as a single number — one dimension, or equal precision in every direction, i.e. \emph{isotropic} precisions, which is the case this paper's examples use — a single attractive cue can pull the report toward another but never \emph{past} it. Eigenvalues in [0, 1] do not give that on their own. With direction-dependent precisions the report can land outside the cues' \emph{convex hull} — the region bounded by the cues themselves, so that landing outside it means landing somewhere no weighted average of the cues could reach — while every eigenvalue of $W$ lies strictly inside (0, 1): $W = \operatorname{diag}(0.9, 0.1)$ sends cues at $(1, 0)$ and $(0, 1)$ to $(0.9, 0.9)$ (companion §8; \texttt{math\_checks.py / cue\_fusion\_commuting\_check}) — a point that is not on the segment joining the two cues at all, so no single blending weight could have produced it. The causal-inference stage (companion §8, Figure 6; Körding et al., 2007) softens this only to \emph{release}, because at large discrepancy the report decays back to the unimodal estimate. In that same scalar case it too is convex-hull-bounded — the report stays between the cues — so it does not by itself produce true \emph{repulsion}. Static Bayesian integration therefore captures only the \emph{congruent, small-discrepancy} regime of these illusions. It is insufficient for adversarial or anatomically impossible conflicts, where genuine \textbf{repulsion} appears. Two such cases fall just outside the model. Adaptation aftereffects are recovered by a Bayesian observer once its encoding is shaped by \textbf{efficient coding} — the principle that a sensory system allocates its limited resolution in proportion to how often each stimulus value actually occurs. Wei and Stocker (2015) show that this allocation makes the likelihood function asymmetric, its heavier tail pointing away from the peak of the prior. The resulting \emph{likelihood repulsion} can outweigh the usual pull toward the prior, so the optimal estimate is biased away from the prior, not toward it. And Tsakiris and Haggard (2005) attribute the implausible-posture sign-reversal of the rubber-hand illusion to a body-schema constraint on the reference frame. The point is scoping, not repair: the reporting node is an integration model, and repulsion is a distinct phenomenon it does not claim.

The most discriminating untested prediction \emph{within pain} is the bifurcation's: that finite-window \textbf{migration declines as sensitization carries the system past its transition} (companion §7.1). Even here the shape is not exotic. A change in variance and instability on the approach to a transition is \textbf{critical slowing down}, a generic early-warning signature documented across ecology, climate, and physiology (Scheffer et al., 2009). That makes a change in instability near a transition plausible in some dynamical systems. It is an analogy and a motivation, not a derivation of this paper's spatial prediction, which stays prospective before any pain data exist.

No study has yet run the decisive within-pain test, so what follows is a design and not a result. It would measure two quantities in the same patients: spatial pain-map instability, as day-to-day drift by ecological momentary assessment, and a \emph{graded} sensitization index, by quantitative sensory testing or the Central Sensitization Inventory. Its primary analysis is one-sided: instability \emph{declines} across the high-gain side of the transition, tested against the null of no decline, and carrying \emph{time-varying} daily distress as a covariate so that the decline cannot be an artifact of affective lability rather than of the $s^{*}$ transition. An earlier form of this design tested instead for an inverted U, a significant negative quadratic term. That is no longer the registered shape, and the reason is quantitative rather than presentational: simulating the reduced dynamics bounds the climb into the interior maximum at 2.2\% against a decline past it of at least 29.5\%, so the model does not license a practically detectable intermediate peak (companion §7.1). A quadratic test spends its power on a feature the model says is too small to find. The one-sided decline test is the weaker claim and the one the mathematics supports; it is also the cheaper study, because detecting a one-directional decline needs fewer participants than resolving curvature. The design is preregistered in the supplement.

A targeted search finds no study testing this decline, and the nearest data are less favourable than they first appear. In frozen shoulder (N = 48), pain-drawing extent correlated with the \emph{questionnaire} index of central sensitization (Central Sensitization Inventory, $r_s$ = .36) but with \textbf{none of the three direct psychophysical measures}. Pressure-pain threshold, temporal summation and conditioned pain modulation all returned small \emph{negative} coefficients ($r_s$ = $-$.12, $-$.15, $-$.14; Balasch-Bernat et al., 2022). Three things follow. First, the association that survives is with the \emph{questionnaire}, not with the psychophysics. The questionnaire is itself only loosely coupled to psychophysical measures (pooled |$r$| $\le $ .27, and $r$ = .002 in healthy controls; Neblett et al., 2024), which is the ground for Velasco et al.'s (2024) caution about questionnaire-defined sensitization. Second, extent correlated as strongly with current pain intensity ($r_s$ = .42) and catastrophizing ($r_s$ = .31) as with the sensitization questionnaire. That is the distress confound of the Limitations, visible in Balasch-Bernat et al.'s (2022) published data. Third, and least comfortably for this paper: those three coefficients are rank correlations against \emph{extent}, and extent is the quantity the model predicts to rise monotonically with gain. A rank correlation detects a monotone rise, so the three nulls are not uninformative — they are weak evidence \emph{against} the predicted rise. They are weak because of the cohort (N = 48), the single pain drawing, and the loose coupling between the two sensitization measures, not because of anything about curve shape; adjudicating the direction needs per-participant values and a gain axis that is graded rather than dichotomized. They are not evidence about \emph{instability}, the quantity the design above measures, for which no study has yet reported a coefficient of any sign. One measurement hazard is worth stating in advance: the high-gain side of the transition, where the predicted decline lives, is also where a body-map can \emph{saturate} (pain everywhere), and a saturated map has little room left to \emph{drift}. A naïve instability metric would then read low variability at high gain and mistake a ceiling artifact for the model's predicted decline. The analysis must therefore define spatial instability so it is not confounded by extent (e.g., a drift measure normalized to the current painful area), or the predicted decline cannot be told apart from saturation.

\section{Re-Examination of Prior Quantitative Claims}

A targeted verification found the well-cited high-utility figures uneven. Knee osteoarthritis clinical criteria (sensitivity 89\%, specificity 88\%; Altman et al., 1986) and Complex Regional Pain Syndrome (Budapest clinical criteria, sensitivity .99, specificity .68; Harden et al., 2010), formerly reflex sympathetic dystrophy and causalgia, hold up. Three others do not. The migraine figures of .81/.75 — sensitivity .81 (95\% CI .77–.85) and specificity .75 (95\% CI .64–.84) — come from the ID-Migraine screener (Lipton et al., 2003), not from diagnostic criteria. The specificity of typical, oppressive chest pain for acute infarction is modest, about .58 on the best primary source (Bruyninckx et al., 2008), which falls well below what the “high-utility” framing implies. And the straight-leg raise is sensitive but not specific (pooled sensitivity .91, 95\% CI .82–.94; pooled specificity .26, 95\% CI .16–.38; Devillé et al., 2000), so its rule-in value is overstated. Recomputing internally consistent likelihood ratios flattens the high end. The asymmetry is real but less dramatic than drawn, and the inflation localizes to specificity.

That last detail is what connects this re-examination to the rest of the paper. A gradient of diagnostic utility looks like a single clean axis partly because its high end has been inflated, and the four biases below are the mechanisms of that inflation. They are not a separate complaint about the literature: they explain why one-gradient framing survived, and they are the reason the paper's three mechanisms could stay pooled inside one utility number without anyone noticing. \textbf{Small-study effects and selective dissemination} are a candidate explanation for the inflation, and a weaker one than we first wrote. Failure to publish and selective reporting are documented in diagnostic-accuracy research — of 418 registered test-accuracy studies, 54\% had been published, and among the 153 published studies that had been registered before completion, a third differed from their registration on eligibility, threshold or outcomes (Korevaar et al., 2014). Smaller studies in this literature do tend to report greater accuracy (Song et al., 2002). But the directional claim — that studies reporting stronger performance are preferentially published — is not established: the two cohort studies that tested it directly found no association between reported accuracy and later full publication (Brazzelli et al., 2009; Korevaar et al., 2016). Both follow studies forward from a conference abstract, so what they rule out is bias in the publication process \emph{after} abstract acceptance; neither can speak to bias at submission or acceptance, and Brazzelli's authors say so. The small-study pattern therefore cannot by itself be attributed to publication bias, and this paper does not attribute it. A registered review would separate the two (see Future Directions). Two \emph{study-level} biases point the same way, and they drive the specific figures flagged above more strongly than the reporting problems just described. \textbf{Verification (work-up) bias} inflates sensitivity and specificity: it arises when only patients with a positive index test receive the reference standard. \textbf{Spectrum bias} inflates both when the validation sample is skewed toward clear-cut cases (Whiting et al., 2004). These are not hypothetical: a large methodological study found that diagnostic-accuracy estimates from such designs are empirically inflated — non-representative case–control sampling overstated the diagnostic odds ratio roughly three-fold (relative diagnostic odds ratio 3.0, 95\% CI 2.0–4.5; Lijmer et al., 1999).

The straight-leg raise shows this inflation most clearly, and Devillé et al. (2000) acknowledge it in their own review. Every study they pooled was a surgical case series outside primary care, and discriminative power fell in the studies with more valid designs — with blinded assessment of index test and reference standard, and with a more homogeneous case mix. Those authors conclude that their studies “do not enable a valid evaluation of the diagnostic accuracy of both tests.” So the upper-bound reading of these figures is not an inference this paper imposes on the literature. Devillé and colleagues say it about their own review. The same reasoning applies to the chest-pain estimates.

\section{Positioning and Novelty}

The three mechanisms are individually recognized in the literature (though none of the three has direct support for its \emph{core} prediction: multiplexing is \emph{consistent} with diagnostic-accuracy data, amplification's conditional migration-decline prediction is \emph{open} with its key test designed, and displacement's location hypothesis is \emph{open} with no direct test yet — Table 3), and multiplexing and displacement are routinely paired in referred-pain reviews (Jin et al., 2023). The contribution is the reframing: treating the diagnostic utility of location as one quantity that fails for three epistemically distinct reasons. That reframing is orthogonal to the nociceptive/neuropathic/nociplastic taxonomy (Kosek et al., 2016), which classifies \emph{what kind of pain} a patient has. Clinically, the reframing changes not the distinctness itself, which any experienced physician knows, but its \emph{consequence}. Because the three are distinct inference problems, each is fixed by a different \emph{sampling} strategy rather than by localizing harder. Multiplexing calls for richer sampling of the same map: provocation, to raise the effective rank of $R$. Amplification calls for sampling over \emph{time}: spatial-instability tracking. Displacement splits in two. Its location hypothesis calls for group-stratified measurement of the observation model under a fixed reference diagnosis. Its care-pathway corollary calls for a change at the treatment layer — a systemic or equity intervention once a signal is detected — and not for anything the individual clinician does at the bedside with the detection threshold. A single “try to localize better” prescription is wrong for all three, and the taxonomy tells you which of them to act on. Bayesian and active-inference accounts of pain exist, but they model perception and intensity: a distinct axis. A candidate novel formal object fills that gap. A \emph{targeted (not systematic)} search of the computational-pain literature surfaced no Bayesian or computational model of pain \emph{spatial} localization uncertainty: a posterior over body location given noisy convergent input. The spatial mislocalization model (companion §8) is proposed to fill it; the search cannot exclude non-indexed or adjacent work, and a registered search would be needed to make the novelty claim firm. Its perceptual stage is deliberately borrowed — it is the linear inverse (lead-field) problem of electroencephalographic source localization (e.g., standardized low-resolution tomography; Pascual-Marqui, 2002), and that stage carries no novelty claim. To our knowledge from that same targeted search, the \emph{cascade} built on it has not been formalized before — a claim bounded by the search limits just stated, not a firm priority claim. It frames referred pain as a within-modality spatial ambiguity problem — many deep generators producing similar distributions over the body-surface field, the ambiguity arising from overlapping referral columns rather than from the surface being two-dimensional (companion §4), isolates the proximal$\to $distal projection bias, and dissociates structural from cognitive mislocalization via the reporting stage. The contribution is therefore the synthesis plus this one model; the remaining mechanisms are established and the framing is partially novel.

\section{Limitations}

No parameter in this paper is estimated from patient data. The models are built to establish that three clinically conflated failures are formally distinct objects — a rank-deficient inverse problem, a bifurcation, and a decision with mediation — and to set the scale of the effect each would produce if it were operating. They show that the three are separable in principle. They do not show how large any of them is in a clinic. Two idealizations deserve flagging. First, the model treats provocation tests as rank-raising, but no manoeuvre loads one structure alone, so the inverse problem stays partially ill-posed. The model's honest output is the residual posterior width $\Sigma _{\mathrm{post}}$, which quantifies that leftover uncertainty rather than removing it (companion §4, §8). Second, the information-monotonicity result assumes a \emph{fixed inferential target}, not a stationary channel or independent observations. Where the cause itself drifts, as in the acute-to-centralized transition, the target becomes a trajectory rather than a fixed cause and the problem becomes filtering a time-varying state — exactly the regime the change-point reading of delocalized amplification is meant to capture (companion §5).

Both results share a third idealization. The companion paper models what a patient \emph{reports}. A physical source can be controlled experimentally — the provocation design below does exactly that — but the latent \emph{felt} location cannot be independently observed, so the percept-to-report step stays partly unverifiable. The nearest available bound comes from a domain where the ground truth does exist. Across 106 effect sizes, self-reported digital media use correlates only moderately with device logs, and the measures framed as \emph{problematic} use track those logs even worse (Parry et al., 2021). Pain has no log. That is the favourable case rather than this one, and it is only moderate. It is a methodological warning about self-report measurement error, not a bound on pain-report reliability nor an ordering of which parameters will be estimated least precisely.

One alternative is strong enough to test first: a distress or somatization confound. A single affective factor could plausibly reproduce all three signatures — widespread and migrating pain, diffuse low-back reporting, and group-patterned reporting. (Catastrophizing, fear-avoidance, and attentional bias are secondary targets for the same test.) Each mechanism must therefore show incremental validity over a distress measure, and none yet has on the outcome that matters here, localization. The number of pain sites strongly predicts long-term disability: a tenfold gradient over 14 years, from zero sites to nine or ten (Kamaleri et al., 2009). That gradient comes from the authors' preferred model, which leaves general health and sick leave out, treating them as downstream consequences of pain and not as confounders. Their fully adjusted model, which keeps both in, shows the association nearly vanish outside the top category (odds ratio 3.92, 95\% CI 1.07–14.37, versus 11.69 with them excluded). Distress itself, however, is not what either model sets aside: a validated general-distress instrument, the 20-item General Health Questionnaire, is a covariate in both, and the site-count gradient survives it — significant from five to six sites upward in the authors' preferred model, and still clearing the top category in the fully adjusted one. So the nearest existing test — general distress, a count of pain sites, disability pensioning as the outcome — has been run, and the count survived. No study has tested the version this framework needs: whether spatial extent survives adjustment for a \emph{pain-specific} distress instrument (catastrophizing or fear-avoidance, the secondary targets above) with \emph{localization} rather than disability as the outcome. One prospective cohort finds distress predicting the \emph{spread} of pain (Christensen et al., 2017), so the confound may be causally upstream rather than separable. The insula dissociation gives a neural reason to expect separability in principle. The incremental-validity test remains open, and is for this framework the most important near-term check.

The literature check was targeted, not systematic, so it cannot quantify \textbf{publication bias}. That bias threatens the argument on two fronts. The diagnostic-accuracy figures the argument leans on are exactly the kind that selective publication of positive test-performance studies, verification bias, and spectrum bias inflate, and the specificity inflation this re-examination turned up looks like exactly that. And the novelty claim rests on the \emph{absence} of an indexed computational model, where absence of evidence is not evidence of absence. For delocalized amplification, even the quantitative sensory testing standard is itself a surrogate for a contested entity.

Mechanisms co-occur, and three boundary cases mark the framework's edges, not its support. Naming them makes the scope explicit. Innocuous warm and cool bars interleaved, felt as burning, are the thermal grill illusion. The illusion is a failure of \emph{quality}, not of location, and reads better as central disinhibition (Craig \& Bushnell, 1994). It is a cousin of delocalized amplification's gain change, and not something the spatial reporting node addresses. Congenital insensitivity to pain from loss-of-function $SCN9A$/Na\_v1.7 mutations (Cox et al., 2006) falls outside the model entirely, because with the nociceptive channel absent there is no signal to localize. Peripheral polyneuropathy, the stocking-glove pain of diabetes, degrades the transmission line itself as cutaneous afferents die. Model E captures its \emph{effect} as an inflated transmission noise $\Sigma _c$ that blurs localization, but it is not one of the three named mechanisms. This account does not claim the three mechanisms exhaust \emph{every} way location can fail. They are the three that the “single gradient” account conflates, not a partition of all of pain.

Those boundary cases mark the edges. Some presentations engage all three failures at once, and those are the natural test-bed. Temporomandibular disorder — pain in the jaw joint and chewing muscles, often with a history of stress and clenching — is the clearest. Its location cannot be read off the map: the trigeminal nerve pools signals from joint, muscle, tooth, and ear, so pain from one is felt in another. That is anatomical multiplexing in the face. The disorder also belongs to the central-sensitization family, one of a cluster of \emph{overlapping chronic pain conditions} (Maixner et al., 2016). Chronic cases show heightened experimental pain sensitivity. But the prospective OPPERA cohort (Orofacial Pain: Prospective Evaluation and Risk Assessment) found that this sensitivity \emph{tracks} the disorder's onset and recovery rather than predicting who develops it. The strongest predictors of onset were instead comorbid conditions and the frequency of somatic symptoms (Slade et al., 2016). That is the distress confound made concrete: a central-amplification signature that is as much consequence as cause. And the disorder displaces. Pain refers to the ear and teeth. In \emph{persistent idiopathic dentoalveolar pain} (formerly atypical odontalgia), pain is placed confidently in a tooth that examination and imaging find normal, and patients undergo repeated dental procedures before it is recognized as having no local cause (Coulter \& Nixdorf, 2020) — a case in which reported location and locally demonstrable pathology diverge. Which node produced that divergence is not settled by the normal work-up: structural referral, a nonlocal nociceptive source, a neuropathic mechanism, a central spatial representation and a distorted report all survive it, and separating perceptual displacement from reporting distortion needs the zero-expectation control proposed for the reporting node. The same clenching history may accompany both a peripheral generator and a central one, which is why the jaw is where the distress confound is hardest to escape. It is also where testing it pays best: OPPERA is prospective and equipped with quantitative sensory testing, a better-resourced setting than low-back pain for the incremental-validity study the framework most needs.

Two commitments bound how this framework may be used. Group-aware criteria are proposed only where the presentation difference has a mechanistic basis, such as biological sex or age. They must not be extended to race or ethnicity as biological categories. That limit follows the ongoing removal of race-based clinical adjustments: the race-free estimated glomerular filtration rate (Delgado et al., 2021) and race-neutral pulmonary-function interpretation (American Thoracic Society, 2023). The second guard concerns the reporting node. Naming a cognitive-expectation stage (Model E's $W$) says \emph{where} location information is distorted. It is not a licence to reattribute pain to the patient's psychology. That failure mode has historically dismissed the suffering of women and minoritized patients as “all in the head.” The framework points the other way. The expectation term formalizes a mechanism \emph{external} to the nociceptive signal, which remains real and intact, and displacement puts the paradigmatic injustice, the missed infarction in women, at a \emph{systemic} threshold and not in the patient. Only a zero-expectation control establishes an expectation effect (companion §8); demographics never do, and neither does a negative work-up. A negative work-up is evidence that the source is not where location suggested, not that the pain is imagined.

\section{Future Directions}

Three studies follow directly. \textbf{Study 1} (amplification) tracks patients' pain maps by phone over several weeks, using ecological momentary assessment (EMA). It tests whether \emph{spatial instability}, meaning how much the felt location drifts from day to day, predicts the outcome (prognosis or treatment response) over and above a single static snapshot and the patient's distress. That distress term must be \emph{time-varying}: daily affect logged in the same EMA instead of a baseline trait score. Otherwise day-to-day emotional lability could manufacture the very spatial instability the test would credit to sensitization, and adjusting only for trait distress leaves the result confoundable. A targeted search surfaced no study of this exact question, variability in \emph{location} rather than in intensity (cf. Pagé et al., 2022). A power analysis (Appendix A) puts the sample needed at roughly 470–590 participants over two to four weeks (for an incremental partial correlation of $\approx $ .15). That is the \emph{detection} target; the \emph{recovery} requirement below is the one that governs the study, at $\approx $ 1250. Two things inflate the detection number. An instability estimate built from 14 to 28 days of sampling is itself noisy, and correcting for that unreliability is most of the cost: the unattenuated requirement is $\approx $ 340 at the same effect size. Widening the plausible effect to .12–.15 gives $\approx $ 340–540 unattenuated; carrying the same reliability correction through at .12 raises the operative requirement to $\approx $ 730–930 (14 to 28 days), so an effect materially smaller than .15 is not adequately powered at that size.

Detecting the association is not the binding constraint. The preregistered decision rule has to tell four explanations apart: the framework's own, a distress marker, a confound, and a mediated effect. Telling them apart is harder than detecting any one of them. Simulating the rule against data generated from each explanation in turn recovers the right one in 88.3\% of runs (the mediated effect is the hardest of the four to tell apart) only at about 1250 participants, two and a half times the detection requirement. That simulation also changed the rule, and the sequence that follows is progressive repair rather than instability: the rule's false-support rate fell from 23\% as first written, to 5.3\% after the standard error was corrected, to 0.22\% once negligibility-first and standardized cross-lag checks were added. As first written it declared the framework supported in about 23\% of runs (Monte Carlo standard error 1.3 points, at 500 participants, a 28-day window and distress reliability 0.85) generated with \textbf{no} path from spatial extent to outcome. It compared the two temporal orderings without first asking whether either differed from zero. Both the repair and the larger sample are registered.

The rule's own uncertainty calculation was then found to be wrong in turn, and rebuilt: it had attached the standard error of an ordinary correlation to a corrected one, understating it by 11\% and running a nominally 5\% test at 8\%. Rebuilt and remeasured, the rule announced the framework supported in 5.3\% of runs where no such path exists. A further repair — checking negligibility before presence, and standardising the cross-lag rather than using it raw — cut that rate to 0.22\% of runs, 95\% interval 0.16\% to 0.29\%, its current stated rate. The 1250-participant figure survives both repairs unchanged. What the repairs do not remove is a dependence the figure carries from the start: 1250 participants suffices if the true association is about .33, but takes roughly 2000 if it is .29, and no sample size tested lets the rule discriminate at .24. Preregistration §7d states the three defects that remain open in the rule and what each would cost to fix.

Continuous body-map EMA is a \emph{research} instrument, not a routine bedside test. It needs app infrastructure and patient adherence, so a positive finding would then have to be distilled into a lightweight clinical proxy, a few well-spaced pain drawings, before it could change practice.

\textbf{Study 2} (displacement) has two parts, aimed at the two separate claims in mechanism (c). It first builds and independently checks group-specific presentation criteria (sex first, then age) — the direct test the location-displacement hypothesis still needs (Table 3) — and, because the care-pathway corollary already shows fixing detection alone did not improve outcomes (Lee et al., 2019), separately tests the \emph{treatment} step itself, likely an implementation trial. \textbf{Study 3} (multiplexing) tests whether a smartphone activity diary of aggravating movements outperforms a single clinic history at localizing the low-back source, quantifying how much of the single-visit floor is recoverable by better sampling.

Beyond these, the clearest theoretical extension is the reporting-stage fourth mechanism, distortion by expectation and by the measuring instrument. The spatial Bayesian model (companion §8) is \emph{estimable}, meaning its parameters can be recovered from data under explicit design conditions: the provocation must keep enough information for stable estimation, the cue locations must be separately anchored, and the design needs at least two trials at distinct, known report weights — a zero-expectation trial being one sufficient way to get the second. Inject hypertonic saline at blinded lumbar levels to drive a known deep source, use a sham “diagnostic laser” to set the patient's expectation, and record where the pain is felt on an iPad body-avatar (ecological momentary assessment). A zero-expectation control trial separates the fixed anatomical shift from the pull of expectation. The simulation flags the hard case to target: telling apart two structures in the \emph{same} spinal segment (the “blobs” risk). The design rule follows from treating provocation as an \emph{instrument-like perturbation} — a designed intervention intended to move one source likelihood and not the others — so a good provocation stresses one structure as selectively as it can. Calling it an instrumental variable in the formal sense would require the exclusion and independence assumptions that go with that term, and the Limitations above say no real manoeuvre loads one structure alone. A registered systematic review would convert the verification pass into a formal evidence appraisal. It would assess publication bias explicitly, through funnel-plot asymmetry and Egger or Peters regression tests where enough studies permit, risk-of-bias grading with QUADAS-2 (Quality Assessment of Diagnostic Accuracy Studies), and certainty ratings with GRADE (Grading of Recommendations, Assessment, Development and Evaluation). That matters because the diagnostic-accuracy literature this account leans on is susceptible to selective publication.

Eight formal extensions would carry these illustrative proofs toward a calibratable instrument. A companion note in the public supplement (\texttt{formal-extensions.md}) develops them with the notation harmonized. Each is a \emph{proposed} formalization, estimable only with the longitudinal data above; none is a fitted result.

Extensions 1 and 2 are the pair to build first: Extension 1 supplies the dynamic state-space substrate, and Extension 2 places a sequential decision rule on top of it. Extension 1 relaxes the fixed-target idealization behind the information chain rule (companion §5): the generator and the referral field co-evolve as a diffusion, and a filter tracks both from irregularly spaced reports. Its fitted widening rate then reads out the onset of delocalized amplification. Extension 2 carries the care-pathway corollary's one-shot decision (companion §6) forward in time as a partially observable Markov decision process, a rule that acts on a belief, not on a known state. It reuses the false-negative and false-positive costs of companion §6. Its policy returns the stopping time at which tracking should give way to treatment. A clinician reads that policy off a table computed offline, rather than solving it at the bedside.

Extensions 3–8 vary that dynamic state-space substrate in six directions. Extension 3 moves the bifurcation (companion §7) onto a connectivity graph, where the instability becomes a closing spectral gap, the shrinking margin by which the slowest mode outlasts the rest. Extension 4 lets the expectation prior $\mu _a$ vary with covariates, splitting reporting error into an anatomical and an expectation component. That split is identified only against the zero-expectation control of companion §8, and a fitted covariate loading measures population-level inequity, never an individual patient's credibility. Extension 5 bounds what a report can carry at all, treating it as a bandwidth-limited channel — a ceiling on every tracker and decision rule downstream, so that ceiling should size the instrument before the rest are fit. Extension 6 gives the tracking dynamics power-law, fractional-order memory, whose fitted order $\alpha $ indexes chronicity. Extension 7 recasts persistent pain as a rigid, high-precision prior that the body then acts to satisfy — a speculative conjecture, not a unification of the three mechanisms, which this framework keeps distinct. Extension 8 models the encounter as a signaling game, in which uninformative reporting is a property of the system's incentives, not of the patient.

The practical benefit is that naming the mechanism changes the target. Where location fails by anatomical multiplexing (a), the remedy is better sampling: provocation and dynamic assessment, not more static imaging. Where it fails by delocalized amplification (b), and if the dynamic prediction is supported, change in the map rather than a single snapshot would be the discriminating signal. Whether body-representation interventions alter the spatial phenotype and clinical outcomes is then a testable therapeutic hypothesis. Interventions that retrain localization and body representation, such as sensorimotor and graded-motor-imagery approaches, are the natural candidates for that test rather than a consequence of the model; their effect sizes remain to be established robustly. Where it fails by structured displacement (c), what has to change is the treatment layer, not only the detection threshold. A single “diagnostic utility” number cannot distinguish these; the framework proposes which of three different things to measure next. If prospective studies validate the mechanism-specific predictions, the same classification would also guide which measurement or intervention to try next.

\section{Conclusion}

The low diagnostic utility of pain location is not one problem but three, with different mathematics, different remedies, and different limits. When many structures share one location, the problem is identifiability, and the model predicts that provocation adds more discriminating information than a finer static localization. If the generator has moved centrally, the problem is that the target itself will not hold still, and the framework motivates mapping it over time. Whether location itself shifts systematically by patient group is a third, still-open question; what is already shown is a downstream, related problem — that better detection alone does not improve outcomes when the bottleneck is treatment, not detection. The two are not the same claim: one concerns the spatial observation channel, the other the downstream care pathway. A fourth node follows the three perceptual mechanisms: the report, which can misplace pain that was felt correctly, and which the companion paper models explicitly — to our knowledge for the first time, a claim held to the limits stated under Positioning and Novelty.

Naming which mechanism is operating turns a vague number into a specific question about \emph{why} location fails, and, for the dynamic mechanism, points to watching how pain moves rather than only where it sits. This paper does not yet have a fitted model: every result here is an existence proof, showing a mechanism \emph{can} produce the observed pattern, not that it does. The preregistration and the three studies above exist to close that gap, and the predictions are stated sharply enough to lose: the framework fails if a single latent axis of spatial precision explains the spatial data as well as the mechanism-specific models do, and if source perturbation, temporal sampling and conditional stratification generate no distinct predictive signatures.

\section{Practical Implications}

\begin{itemize}\setlength{\itemsep}{2pt}\item \textbf{Before asking how useful location is, ask which failure you are in.} The answer changes the instrument, not merely the confidence attached to the utility number.\item \textbf{When several structures share a segment, the framework predicts more is gained by loading the candidates than by refining the localization question} — and by provocation tests in combination, since single manual tests are uninformative in isolation.\item \textbf{When extent or location is the outcome, measure it more than once.} A single pain drawing cannot detect drift, and under this account any drift is a candidate signal; whether it is the predicted displacement mode is still an empirical question.\item \textbf{Report diagnostic accuracy stratified by the covariate, not pooled.} A pooled figure can conceal poor performance in one or more groups and hide materially different error profiles.\item \textbf{Do not treat improved detection as the outcome.} Where the treatment step lags for a group, detection gains do not transfer to that group.\end{itemize}

And four pitfalls, which are the same claims stated as the errors they prevent.

\begin{itemize}\setlength{\itemsep}{2pt}\item \textbf{Do not read a pooled diagnostic-utility figure as a property of the symptom.} It is a property of the symptom \emph{in that population under that mechanism}, and the three mechanisms carry different values.\item \textbf{Do not read a null correlation between extent and sensitization as neutral about this model.} Extent is the order parameter and is predicted to rise monotonically with gain, a shape a rank correlation can detect, so a null is weak evidence \emph{against} the prediction. The published summaries disagree in sign (Balasch-Bernat et al., 2022; Fernández-de-las-Peñas et al., 2022), which is the firmer finding and the reason no single linear summary of extent travels between populations.\item \textbf{Do not infer central amplification from a questionnaire alone.} Its association with psychophysical measures is weak (no modality exceeds |$r$| = .27, and the pressure-threshold association disappears entirely in healthy controls), and the construct itself is contested.\item \textbf{Do not adjust for distress with a baseline trait score when the outcome is spatial instability.} Emotional lability varies day to day and can manufacture the instability being measured, so the distress term has to be time-varying or the result stays confoundable.\end{itemize}

\section{Acknowledgments}

The author thanks J.S. and C.S. for critical review of the manuscript, and colleagues and readers for discussion of these ideas.

\section{Statements and Declarations}

\textbf{Funding.} The author declares that no funds, grants, or other support were received during the preparation of this manuscript.

\textbf{Competing interests.} The author has no relevant financial or non-financial interests to disclose.

\textbf{Ethics approval.} Not required. This is a conceptual and methodological Perspective; it reports no study involving human participants or animals, and analyses no individual-level human data. All quantitative results are simulations from the models described in the companion paper, run on synthetic inputs, together with re-analysis of summary statistics already published elsewhere.

\textbf{Consent to participate / consent to publish.} Not applicable, for the reason above.

\textbf{Author contributions.} A.Y.S. is the sole author. He conceived the framework, specified and implemented every model, performed all analyses, verified the cited quantitative claims against their primary sources, and wrote and revised the manuscript. Assistance from AI tools, and the limits placed on it, are disclosed in full below.

\textbf{Use of AI tools.} AI tools assisted under the author's direction. Claude (Anthropic) helped implement the simulation code and figures, gave editorial support, and helped organise the argument for three audiences. Perplexity was used for literature search and citation verification. Gemini (Google) and ChatGPT (OpenAI) were used for adversarial review and independent mathematical checking; ChatGPT's review of the mathematics prompted the claim-status audit in companion §10.

No claim rests on an AI's word. Every quantitative claim here was checked against its primary source. For one source that check reached the published abstract rather than the full text, and it is named here: Brazzelli et al. (2009). Its abstract is verified verbatim, including the null result the argument turns on, and the published paper is on file, but its body has not been read. The verification trail is retained in the supplement: a per-reference registry records the class of evidence behind every source, and the reading ledger names every source whose full text remains unread. That trail also records what the checking caught — AI-supplied citations, since excluded; four claims that misstated their source; a reference-list entry spliced together from two others; and one figure the author had wrongly \emph{removed} on a bad inference. Several were caught after the preprint was posted. The process is still running, which is why it is documented.

The author is solely accountable for the content, including all errors. Consistent with ICMJE and COPE guidance, no AI system is an author or can hold accountability for the work.

\section{Data and Code Availability}

Every model, figure, and sensitivity sweep in this paper regenerates from the public code archive on the Open Science Framework (\url{https://osf.io/zfhkq/}), persistent identifier \url{https://doi.org/10.17605/OSF.IO/ZFHKQ}. The archive also holds the preregistration of Study 1, the model-testing registry that Table 3 summarizes, and the provenance trail for each verified citation. The study uses \textbf{no primary human-subjects data}: the models are illustrative and parameterized as described.

\section{References}

\par\hangindent=1.5em\noindent Adamczyk, W. M., Saulicz, O., Saulicz, E., \& Luedtke, K. (2018). Tactile acuity (dys)function in acute nociceptive low back pain: A double-blind experiment. \emph{Pain, 159}(3), 427–436. https://doi.org/10.1097/j.pain.0000000000001110

\par\hangindent=1.5em\noindent Adamczyk, W., Luedtke, K., \& Saulicz, E. (2018). Lumbar tactile acuity in patients with low back pain and healthy controls: Systematic review and meta-analysis. \emph{The Clinical Journal of Pain, 34}(1), 82–94. https://doi.org/10.1097/AJP.0000000000000499

\par\hangindent=1.5em\noindent Alais, D., \& Burr, D. (2004). The ventriloquist effect results from near-optimal bimodal integration. \emph{Current Biology, 14}(3), 257–262. https://doi.org/10.1016/j.cub.2004.01.029

\par\hangindent=1.5em\noindent Altman, R., Asch, E., Bloch, D., Bole, G., Borenstein, D., Brandt, K., Christy, W., Cooke, T. D. V., Greenwald, R., Hochberg, M., Howell, D. S., Kaplan, D. L., Koopman, W. J., Longley, S., Mankin, H. J., McShane, D. J., Medsger, T. A., Jr., Meenan, R. F., Mikkelsen, W. M., . . . Wolfe, F. (1986). Development of criteria for the classification and reporting of osteoarthritis: Classification of osteoarthritis of the knee. \emph{Arthritis \& Rheumatism, 29}(8), 1039–1049. https://doi.org/10.1002/art.1780290816

\par\hangindent=1.5em\noindent Amari, S. (1977). Dynamics of pattern formation in lateral-inhibition type neural fields. \emph{Biological Cybernetics, 27}(2), 77–87. https://doi.org/10.1007/BF00337259

\par\hangindent=1.5em\noindent American Thoracic Society. (2023). Race and ethnicity in pulmonary function test interpretation: An official American Thoracic Society statement. \emph{American Journal of Respiratory and Critical Care Medicine, 207}(8), 978–995. https://doi.org/10.1164/rccm.202302-0310ST

\par\hangindent=1.5em\noindent Balasch-Bernat, M., Dueñas, L., Aguilar-Rodríguez, M., Falla, D., Schneebeli, A., Navarro-Bosch, M., Lluch, E., \& Barbero, M. (2022). The spatial extent of pain is associated with pain intensity, catastrophizing and some measures of central sensitization in people with frozen shoulder. \emph{Journal of Clinical Medicine, 11}(1), 154. https://doi.org/10.3390/jcm11010154

\par\hangindent=1.5em\noindent Brazzelli, M., Lewis, S. C., Deeks, J. J., \& Sandercock, P. A. G. (2009). No evidence of bias in the process of publication of diagnostic accuracy studies in stroke submitted as abstracts. \emph{Journal of Clinical Epidemiology, 62}(4), 425–430. https://doi.org/10.1016/j.jclinepi.2008.06.018

\par\hangindent=1.5em\noindent Bruyninckx, R., Aertgeerts, B., Bruyninckx, P., \& Buntinx, F. (2008). Signs and symptoms in diagnosing acute myocardial infarction and acute coronary syndrome: A diagnostic meta-analysis. \emph{British Journal of General Practice, 58}(547), e1–e8. https://doi.org/10.3399/bjgp08X277014

\par\hangindent=1.5em\noindent Caseiro, M., Reis, F. J. J. dos, Barbosa, A. M., Barbero, M., Falla, D., \& Oliveira, A. S. de (2021). Two-point discrimination and judgment of laterality in individuals with chronic unilateral non-traumatic shoulder pain. \emph{Musculoskeletal Science and Practice, 56}, 102447. https://doi.org/10.1016/j.msksp.2021.102447

\par\hangindent=1.5em\noindent Catley, M. J., O'Connell, N. E., Berryman, C., Ayhan, F. F., \& Moseley, G. L. (2014). Is tactile acuity altered in people with chronic pain? A systematic review and meta-analysis. \emph{The Journal of Pain, 15}(10), 985–1000. https://doi.org/10.1016/j.jpain.2014.06.009

\par\hangindent=1.5em\noindent Chan, B. L., Witt, R., Charrow, A. P., Magee, A., Howard, R., Pasquina, P. F., Heilman, K. M., \& Tsao, J. W. (2007). Mirror therapy for phantom limb pain. \emph{The New England Journal of Medicine, 357}(21), 2206–2207. https://doi.org/10.1056/NEJMc071927

\par\hangindent=1.5em\noindent Chou, R., Fu, R., Carrino, J. A., \& Deyo, R. A. (2009). Imaging strategies for low-back pain: Systematic review and meta-analysis. \emph{The Lancet, 373}(9662), 463–472. https://doi.org/10.1016/S0140-6736(09)60172-0

\par\hangindent=1.5em\noindent Christensen, J. O., Johansen, S., \& Knardahl, S. (2017). Psychological predictors of change in the number of musculoskeletal pain sites among Norwegian employees: A prospective study. \emph{BMC Musculoskeletal Disorders, 18}(1), 140. https://doi.org/10.1186/s12891-017-1503-7

\par\hangindent=1.5em\noindent Cohen, J. (1988). \emph{Statistical power analysis for the behavioral sciences} (2nd ed.). Lawrence Erlbaum Associates.

\par\hangindent=1.5em\noindent Coulter, J., \& Nixdorf, D. R. (2020). A review of persistent idiopathic dentoalveolar pain (formerly PDAP/atypical odontalgia). \emph{Oral Surgery, 13}(4), 371–378. https://doi.org/10.1111/ors.12472

\par\hangindent=1.5em\noindent Cox, J. J., Reimann, F., Nicholas, A. K., Thornton, G., Roberts, E., Springell, K., Karbani, G., Jafri, H., Mannan, J., Raashid, Y., Al-Gazali, L., Hamamy, H., Valente, E. M., Gorman, S., Williams, R., McHale, D. P., Wood, J. N., Gribble, F. M., \& Woods, C. G. (2006). An SCN9A channelopathy causes congenital inability to experience pain. \emph{Nature, 444}(7121), 894–898. https://doi.org/10.1038/nature05413

\par\hangindent=1.5em\noindent Craig, A. D., \& Bushnell, M. C. (1994). The thermal grill illusion: Unmasking the burn of cold pain. \emph{Science, 265}(5169), 252–255. https://doi.org/10.1126/science.8023144

\par\hangindent=1.5em\noindent Delgado, C., Baweja, M., Crews, D. C., Eneanya, N. D., Gadegbeku, C. A., Inker, L. A., Mendu, M. L., Miller, W. G., Moxey-Mims, M. M., Roberts, G. V., St Peter, W. L., Warfield, C., \& Powe, N. R. (2021). A unifying approach for GFR estimation: Recommendations of the NKF-ASN Task Force on reassessing the inclusion of race in diagnosing kidney disease. \emph{Journal of the American Society of Nephrology, 32}(12), 2994–3015. https://doi.org/10.1681/ASN.2021070988

\par\hangindent=1.5em\noindent Devillé, W. L. J. M., van der Windt, D. A. W. M., Dzaferagić, A., Bezemer, P. D., \& Bouter, L. M. (2000). The test of Lasègue: Systematic review of the accuracy in diagnosing herniated discs. \emph{Spine, 25}(9), 1140–1147. https://doi.org/10.1097/00007632-200005010-00016

\par\hangindent=1.5em\noindent Ehrenbrusthoff, K., Ryan, C. G., Grüneberg, C., \& Martin, D. J. (2018). A systematic review and meta-analysis of the reliability and validity of sensorimotor measurement instruments in people with chronic low back pain. \emph{Musculoskeletal Science and Practice, 35}, 73–83. https://doi.org/10.1016/j.msksp.2018.02.007

\par\hangindent=1.5em\noindent Ernst, M. O., \& Banks, M. S. (2002). Humans integrate visual and haptic information in a statistically optimal fashion. \emph{Nature, 415}(6870), 429–433. https://doi.org/10.1038/415429a

\par\hangindent=1.5em\noindent Ernst, M. O., \& Bülthoff, H. H. (2004). Merging the senses into a robust percept. \emph{Trends in Cognitive Sciences, 8}(4), 162–169. https://doi.org/10.1016/j.tics.2004.02.002

\par\hangindent=1.5em\noindent Fernández-de-las-Peñas, C., Fuensalida-Novo, S., Ortega-Santiago, R., Valera-Calero, J. A., Cescon, C., Derboni, M., Giuffrida, V., \& Barbero, M. (2022). Pain extent is not associated with sensory-associated symptoms, cognitive or psychological variables in COVID-19 survivors suffering from post-COVID pain. \emph{Journal of Clinical Medicine, 11}(15), 4633. https://doi.org/10.3390/jcm11154633

\par\hangindent=1.5em\noindent Gebhart, G. F., \& Bielefeldt, K. (2016). Physiology of visceral pain. \emph{Comprehensive Physiology, 6}(4), 1609–1633. https://doi.org/10.1002/cphy.c150049

\par\hangindent=1.5em\noindent Goldreich, D. (2007). A Bayesian perceptual model replicates the cutaneous rabbit and other tactile spatiotemporal illusions. \emph{PLoS ONE, 2}(3), e333. https://doi.org/10.1371/journal.pone.0000333

\par\hangindent=1.5em\noindent Graven-Nielsen, T., Arendt-Nielsen, L., Svensson, P., \& Jensen, T. S. (1997). Quantification of local and referred muscle pain in humans after sequential i.m. injections of hypertonic saline. \emph{Pain, 69}(1–2), 111–117. https://doi.org/10.1016/S0304-3959(96)03243-5

\par\hangindent=1.5em\noindent Han, C. S., Hancock, M. J., Sharma, S., Sharma, S., Harris, I. A., Cohen, S. P., Magnussen, J., Maher, C. G., \& Traeger, A. C. (2023). Low back pain of disc, sacroiliac joint, or facet joint origin: A diagnostic accuracy systematic review. \emph{eClinicalMedicine, 59}, 101960. https://doi.org/10.1016/j.eclinm.2023.101960

\par\hangindent=1.5em\noindent Hancock, M. J., Maher, C. G., Latimer, J., Spindler, M. F., McAuley, J. H., Laslett, M., \& Bogduk, N. (2007). Systematic review of tests to identify the disc, SIJ or facet joint as the source of low back pain. \emph{European Spine Journal, 16}(10), 1539–1550. https://doi.org/10.1007/s00586-007-0391-1

\par\hangindent=1.5em\noindent Harden, R. N., Bruehl, S., Perez, R. S. G. M., Birklein, F., Marinus, J., Maihofner, C., Lubenow, T., Buvanendran, A., Mackey, S., Graciosa, J., Mogilevski, M., Ramsden, C., Chont, M., \& Vatine, J.-J. (2010). Validation of proposed diagnostic criteria (the "Budapest Criteria") for complex regional pain syndrome. \emph{Pain, 150}(2), 268–274. https://doi.org/10.1016/j.pain.2010.04.030

\par\hangindent=1.5em\noindent Hong, S. W., Xu, L., Kang, M.-S., \& Tong, F. (2012). The hand-reversal illusion revisited. \emph{Frontiers in Integrative Neuroscience, 6}, 83. https://doi.org/10.3389/fnint.2012.00083

\par\hangindent=1.5em\noindent Jarvik, J. G., Hollingworth, W., Martin, B., Emerson, S. S., Gray, D. T., Overman, S., Robinson, D., Staiger, T., Wessbecher, F., Sullivan, S. D., Kreuter, W., \& Deyo, R. A. (2003). Rapid magnetic resonance imaging vs radiographs for patients with low back pain: A randomized controlled trial. \emph{JAMA, 289}(21), 2810–2818. https://doi.org/10.1001/jama.289.21.2810

\par\hangindent=1.5em\noindent Jensen, M. C., Brant-Zawadzki, M. N., Obuchowski, N., Modic, M. T., Malkasian, D., \& Ross, J. S. (1994). Magnetic resonance imaging of the lumbar spine in people without back pain. \emph{The New England Journal of Medicine, 331}(2), 69–73. https://doi.org/10.1056/NEJM199407143310201

\par\hangindent=1.5em\noindent Jin, Q., Chang, Y., Lu, C., Chen, L., \& Wang, Y. (2023). Referred pain: Characteristics, possible mechanisms, and clinical management. \emph{Frontiers in Neurology, 14}, 1104817. https://doi.org/10.3389/fneur.2023.1104817

\par\hangindent=1.5em\noindent Jung, W.-M., Lee, I.-S., Lee, Y.-S., Kim, J., Park, H.-J., Wallraven, C., \& Chae, Y. (2019). Decoding spatial location of perceived pain to acupuncture needle using multivoxel pattern analysis. \emph{Molecular Pain, 15}. https://doi.org/10.1177/1744806919877060

\par\hangindent=1.5em\noindent Kamaleri, Y., Natvig, B., Ihlebæk, C. M., \& Bruusgaard, D. (2009). Does the number of musculoskeletal pain sites predict work disability? A 14-year prospective study. \emph{European Journal of Pain, 13}(4), 426–430. https://doi.org/10.1016/j.ejpain.2008.05.009

\par\hangindent=1.5em\noindent Korevaar, D. A., Ochodo, E. A., Bossuyt, P. M. M., \& Hooft, L. (2014). Publication and reporting of test accuracy studies registered in ClinicalTrials.gov. \emph{Clinical Chemistry, 60}(4), 651–659. https://doi.org/10.1373/clinchem.2013.218149

\par\hangindent=1.5em\noindent Korevaar, D. A., Cohen, J. F., Spijker, R., Saldanha, I. J., Dickersin, K., Virgili, G., Hooft, L., \& Bossuyt, P. M. M. (2016). Reported estimates of diagnostic accuracy in ophthalmology conference abstracts were not associated with full-text publication. \emph{Journal of Clinical Epidemiology, 79}, 96–103. https://doi.org/10.1016/j.jclinepi.2016.06.002

\par\hangindent=1.5em\noindent Kosek, E., Cohen, M., Baron, R., Gebhart, G. F., Mico, J.-A., Rice, A. S. C., Rief, W., \& Sluka, A. K. (2016). Do we need a third mechanistic descriptor for chronic pain states? \emph{Pain, 157}(7), 1382–1386. https://doi.org/10.1097/j.pain.0000000000000507

\par\hangindent=1.5em\noindent Kurosawa, D., Murakami, E., \& Aizawa, T. (2015). Referred pain location depends on the affected section of the sacroiliac joint. \emph{European Spine Journal, 24}(3), 521–527. https://doi.org/10.1007/s00586-014-3604-4

\par\hangindent=1.5em\noindent Körding, K. P., Beierholm, U., Ma, W. J., Quartz, S., Tenenbaum, J. B., \& Shams, L. (2007). Causal inference in multisensory perception. \emph{PLoS ONE, 2}(9), e943. https://doi.org/10.1371/journal.pone.0000943

\par\hangindent=1.5em\noindent Lee, K. K., Ferry, A. V., Anand, A., Strachan, F. E., Chapman, A. R., Kimenai, D. M., Meex, S. J. R., Berry, C., Findlay, I., Reid, A., Cruickshank, A., Gray, A., Collinson, P. O., Apple, F. S., McAllister, D. A., Maguire, D., Fox, K. A. A., Newby, D. E., Tuck, C., . . . Mills, N. L. (2019). Sex-specific thresholds of high-sensitivity troponin in patients with suspected acute coronary syndrome. \emph{Journal of the American College of Cardiology, 74}(16), 2032–2043. https://doi.org/10.1016/j.jacc.2019.07.082

\par\hangindent=1.5em\noindent Li, Z., Wereski, R., Anand, A., Lowry, M. T. H., Doudesis, D., McDermott, M., Ferry, A. V., Tuck, C., Chapman, A. R., Lee, K. K., Shah, A. S. V., Mills, N. L., \& Kimenai, D. M. (2024). Uniform or sex-specific cardiac troponin thresholds to rule out myocardial infarction at presentation. \emph{Journal of the American College of Cardiology, 83}(19), 1855–1866. https://doi.org/10.1016/j.jacc.2024.03.365

\par\hangindent=1.5em\noindent Lijmer, J. G., Mol, B. W., Heisterkamp, S., Bonsel, G. J., Prins, M. H., van der Meulen, J. H. P., \& Bossuyt, P. M. M. (1999). Empirical evidence of design-related bias in studies of diagnostic tests. \emph{JAMA, 282}(11), 1061–1066. https://doi.org/10.1001/jama.282.11.1061

\par\hangindent=1.5em\noindent Lipton, R. B., Dodick, D., Sadovsky, R., Kolodner, K., Endicott, J., Hettiarachchi, J., \& Harrison, W. (2003). A self-administered screener for migraine in primary care: The ID Migraine validation study. \emph{Neurology, 61}(3), 375–382. https://doi.org/10.1212/01.wnl.0000078940.53438.83

\par\hangindent=1.5em\noindent Lloyd, D. M. (2007). Spatial limits on referred touch to an alien limb may reflect boundaries of visuo-tactile peripersonal space surrounding the hand. \emph{Brain and Cognition, 64}(1), 104–109. https://doi.org/10.1016/j.bandc.2006.09.013

\par\hangindent=1.5em\noindent Lu, C., Yang, T., Zhao, H., Zhang, M., Meng, F., Fu, H., Xie, Y., \& Xu, H. (2016). Insular cortex is critical for the perception, modulation, and chronification of pain. \emph{Neuroscience Bulletin, 32}(2), 191–201. https://doi.org/10.1007/s12264-016-0016-y

\par\hangindent=1.5em\noindent Maher, C., Underwood, M., \& Buchbinder, R. (2017). Non-specific low back pain. \emph{The Lancet, 389}(10070), 736–747. https://doi.org/10.1016/S0140-6736(16)30970-9

\par\hangindent=1.5em\noindent Maixner, W., Fillingim, R. B., Williams, D. A., Smith, S. B., \& Slade, G. D. (2016). Overlapping chronic pain conditions: Implications for diagnosis and classification. \emph{The Journal of Pain, 17}(9, Suppl.), T93–T107. https://doi.org/10.1016/j.jpain.2016.06.002

\par\hangindent=1.5em\noindent Mancini, F., Bauleo, A., Cole, J., Lui, F., Porro, C. A., Haggard, P., \& Iannetti, G. D. (2014). Whole-body mapping of spatial acuity for pain and touch. \emph{Annals of Neurology, 75}(6), 917–924. https://doi.org/10.1002/ana.24179

\par\hangindent=1.5em\noindent May, S., \& Aina, A. (2012). Centralization and directional preference: A systematic review. \emph{Manual Therapy, 17}(6), 497–506. https://doi.org/10.1016/j.math.2012.05.003

\par\hangindent=1.5em\noindent Neblett, R., Sanabria-Mazo, J. P., Luciano, J. V., Mirčić, M., Čolović, P., Bojanić, M., Jeremić-Knežević, M., Aleksandrić, T., \& Knežević, A. (2024). Is the Central Sensitization Inventory (CSI) associated with quantitative sensory testing (QST)? A systematic review and meta-analysis. \emph{Neuroscience \& Biobehavioral Reviews, 161}, 105612. https://doi.org/10.1016/j.neubiorev.2024.105612

\par\hangindent=1.5em\noindent Nijs, J., Lahousse, A., Kapreli, E., Bilika, P., Saraçoğlu, İ., Malfliet, A., Coppieters, I., De Baets, L., Leysen, L., Roose, E., Clark, J., Voogt, L., \& Huysmans, E. (2021). Nociplastic pain criteria or recognition of central sensitization? Pain phenotyping in the past, present and future. \emph{Journal of Clinical Medicine, 10}(15), 3203. https://doi.org/10.3390/jcm10153203

\par\hangindent=1.5em\noindent O'Neill, C. W., Kurgansky, M. E., Derby, R., \& Ryan, D. P. (2002). Disc stimulation and patterns of referred pain. \emph{Spine, 27}(24), 2776–2781. https://doi.org/10.1097/00007632-200212150-00007

\par\hangindent=1.5em\noindent Pagé, M. G., Gauvin, L., Sylvestre, M.-P., Nitulescu, R., Dyachenko, A., \& Choinière, M. (2022). An ecological momentary assessment study of pain intensity variability: Ascertaining extent, predictors, and associations with quality of life, interference and health care utilization among individuals living with chronic low back pain. \emph{The Journal of Pain, 23}(7), 1151–1166. https://doi.org/10.1016/j.jpain.2022.01.001

\par\hangindent=1.5em\noindent Parry, D. A., Davidson, B. I., Sewall, C. J. R., Fisher, J. T., Mieczkowski, H., \& Quintana, D. S. (2021). A systematic review and meta-analysis of discrepancies between logged and self-reported digital media use. \emph{Nature Human Behaviour, 5}(11), 1535–1547. https://doi.org/10.1038/s41562-021-01117-5

\par\hangindent=1.5em\noindent Pascual-Marqui, R. D. (2002). Standardized low-resolution brain electromagnetic tomography (sLORETA): Technical details. \emph{Methods and Findings in Experimental and Clinical Pharmacology, 24}(Suppl D), 5–12.

\par\hangindent=1.5em\noindent Ramachandran, V. S., \& Rogers-Ramachandran, D. (1996). Synaesthesia in phantom limbs induced with mirrors. \emph{Proceedings of the Royal Society B, 263}(1369), 377–386. https://doi.org/10.1098/rspb.1996.0058

\par\hangindent=1.5em\noindent Righini, M., Van Es, J., Den Exter, P. L., Roy, P.-M., Verschuren, F., Ghuysen, A., Rutschmann, O. T., Sanchez, O., Jaffrelot, M., Trinh-Duc, A., Le Gall, C., Moustafa, F., Principe, A., Van Houten, A. A., Ten Wolde, M., Douma, R. A., Hazelaar, G., Erkens, P. M. G., Van Kralingen, K. W., . . . Le Gal, G. (2014). Age-adjusted D-dimer cutoff levels to rule out pulmonary embolism: The ADJUST-PE study. \emph{JAMA, 311}(11), 1117–1124. https://doi.org/10.1001/jama.2014.2135

\par\hangindent=1.5em\noindent Scheffer, M., Bascompte, J., Brock, W. A., Brovkin, V., Carpenter, S. R., Dakos, V., Held, H., van Nes, E. H., Rietkerk, M., \& Sugihara, G. (2009). Early-warning signals for critical transitions. \emph{Nature, 461}(7260), 53–59. https://doi.org/10.1038/nature08227

\par\hangindent=1.5em\noindent Schlereth, T., Magerl, W., \& Treede, R.-D. (2001). Spatial discrimination thresholds for pain and touch in human hairy skin. \emph{Pain, 92}(1–2), 187–194. https://doi.org/10.1016/S0304-3959(00)00484-X

\par\hangindent=1.5em\noindent Shah, A. S. V., Griffiths, M., Lee, K. K., McAllister, D. A., Hunter, A. L., Ferry, A. V., Cruikshank, A., Reid, A., Stoddart, M., Strachan, F., Walker, S., Collinson, P. O., Apple, F. S., Gray, A. J., Fox, K. A. A., Newby, D. E., \& Mills, N. L. (2015). High sensitivity cardiac troponin and the under-diagnosis of myocardial infarction in women: Prospective cohort study. \emph{BMJ, 350}, g7873. https://doi.org/10.1136/bmj.g7873

\par\hangindent=1.5em\noindent Shavit, A. Y. (2026). One inference, four nodes: Formal models of why pain location fails. \emph{arXiv}.

\par\hangindent=1.5em\noindent Slade, G. D., Ohrbach, R., Greenspan, J. D., Fillingim, R. B., Bair, E., Sanders, A. E., Dubner, R., Diatchenko, L., Meloto, C. B., Smith, S., \& Maixner, W. (2016). Painful temporomandibular disorder: Decade of discovery from the OPPERA studies. \emph{Journal of Dental Research, 95}(10), 1084–1092. https://doi.org/10.1177/0022034516653743

\par\hangindent=1.5em\noindent Song, F., Khan, K. S., Dinnes, J., \& Sutton, A. J. (2002). Asymmetric funnel plots and publication bias in meta-analyses of diagnostic accuracy. \emph{International Journal of Epidemiology, 31}(1), 88–95. https://doi.org/10.1093/ije/31.1.88

\par\hangindent=1.5em\noindent Tsakiris, M., \& Haggard, P. (2005). The rubber hand illusion revisited: Visuotactile integration and self-attribution. \emph{Journal of Experimental Psychology: Human Perception and Performance, 31}(1), 80–91. https://doi.org/10.1037/0096-1523.31.1.80

\par\hangindent=1.5em\noindent Van Riper, C. (1935). An experimental study of the Japanese illusion. \emph{The American Journal of Psychology, 47}(2), 252–263.

\par\hangindent=1.5em\noindent Velasco, E., Flores-Cortés, M., Guerra-Armas, J., Flix-Díez, L., Gurdiel-Álvarez, F., Donado-Bermejo, A., van den Broeke, E. N., Pérez-Cervera, L., \& Delicado-Miralles, M. (2024). Is chronic pain caused by central sensitization? A review and critical point of view. \emph{Neuroscience \& Biobehavioral Reviews, 167}, 105886. https://doi.org/10.1016/j.neubiorev.2024.105886

\par\hangindent=1.5em\noindent Wagner, J. M., McKinney, W. P., \& Carpenter, J. L. (1996). Does this patient have appendicitis? \emph{JAMA, 276}(19), 1589–1594. PMID 8918857.

\par\hangindent=1.5em\noindent Wand, B. M., Keeves, J., Bourgoin, C., George, P. J., Smith, A. J., O'Connell, N. E., \& Moseley, G. L. (2013). Mislocalization of sensory information in people with chronic low back pain: A preliminary investigation. \emph{The Clinical Journal of Pain, 29}(8), 737–743. https://doi.org/10.1097/AJP.0b013e318274b320

\par\hangindent=1.5em\noindent Wei, X.-X., \& Stocker, A. A. (2015). A Bayesian observer model constrained by efficient coding can explain "anti-Bayesian" percepts. \emph{Nature Neuroscience, 18}(10), 1509–1517. https://doi.org/10.1038/nn.4105

\par\hangindent=1.5em\noindent Whiting, P., Rutjes, A. W. S., Reitsma, J. B., Glas, A. S., Bossuyt, P. M. M., \& Kleijnen, J. (2004). Sources of variation and bias in studies of diagnostic accuracy: A systematic review. \emph{Annals of Internal Medicine, 140}(3), 189–202. https://doi.org/10.7326/0003-4819-140-3-200402030-00010

\par\hangindent=1.5em\noindent Yeh, B. (2008). Does this adult patient have appendicitis? [Evidence-Based Emergency Medicine / Rational Clinical Examination abstract of Wagner et al., 1996.] \emph{Annals of Emergency Medicine, 52}(3), 301–303. https://doi.org/10.1016/j.annemergmed.2007.10.023

\section*{Appendix A: Closed-form power (Study 1)}

For a single incremental predictor, power follows a non-central $\chi $$^{2}$(1) with non-centrality $\lambda  = N f^{2}$, $f^{2} = r_p^{2}/(1 - r_p^{2})$ (Cohen's f$^{2}$; $r_p$ = partial correlation of instability with the outcome after adjustment; Cohen, 1988). The normal approximation gives $N = (z_{1-\alpha /2} + z_{1-\beta })^{2} / f^{2}$ (= 7.85/f$^{2}$ at $\alpha $ = .05, power = .80). Thus $r_p \approx  .15$ requires $N \approx  341$; measurement reliability $\operatorname{Rel}(T) = T/(T+k)$ attenuates $r_p$ by $\sqrt{\operatorname{Rel}(T)}$, so EMA days saturate by two to four weeks. In plain terms, the required sample rises roughly as the inverse square of the adjusted association, so modest attenuation is expensive; and reliability rises quickly in the first days of sampling and then approaches a ceiling.

\textbf{These figures size the study for \emph{detection}, but recovery sets the sample.} Recovering which of four explanations generated the data is a harder requirement than detecting an association, and it sets the registered sample at $\approx $ 1250. The simulation establishing that is in the preregistration (§7b), and its output ships with the code supplement. It also found and repaired a defect in the decision rule, which returned the framework's own scenario in about 23\% of runs (Monte Carlo standard error 1.3 points, at 500 participants, a 28-day window and distress reliability 0.85) generated with no extent-to-outcome path.

\textbf{The simulation was recomputed on 2026-09-17 and the 1250 stands.} The uncertainty calculation behind it had understated the corrected estimate's standard error by 11\%, so a test reported at 5\% was running at 8\%; rebuilt, it ran at 5.3\% (95\% interval 5.0\% to 5.6\%). A further repair — checking negligibility before presence, and standardising the cross-lag rather than using it raw — cut that rate to 0.22\% (95\% interval 0.16\% to 0.29\%); two independently seeded sweeps put the sample size at the same 1250 under both repairs. Two things about that number did change. At the registered 1250, the \emph{mediated effect} is the harder of the two to recover — 88.3\% against the confound's 89.5\% — but the confound is the scenario whose recovery does not improve with more participants (89.5\% at 1250, 90.2\% at 3000), while the mediated effect's recovery keeps climbing (88.3\% to 99.9\% over that same range). Past roughly 1250–1500 participants the confound overtakes the mediated effect and becomes, and remains, the scenario more participants do not buy. And the figure is conditional on an assumed association of about .33; at .29 it becomes roughly 2000, and at .24 no tested sample size suffices. Preregistration §7d records three defects in the rule that remain open, each with the measured size of its effect.

\section*{Appendix B: Glossary}

Plain definitions of terms that are standard in one field but unfamiliar in another. This complements the table of shared terms in \emph{Reading across three fields}: that table handles words that mean \emph{different} things to each audience; this one defines words one audience will not know. Grouped by the field a term comes from and the readers who will need it.

\section{Clinical \& diagnostic terms (for the mathematics and psychology reader)}

{\footnotesize\begin{xltabular}{\textwidth}{>{\raggedright\arraybackslash}X>{\raggedright\arraybackslash}X}
\toprule
Term & Plain meaning \\
\midrule
\endhead
Likelihood ratio (LR+, LR$-$) & how much a test result multiplies the odds of disease; LR+ > 1 rules in, LR$-$ < 1 rules out \\
Reference standard & the best available “truth” a test is judged against (here, diagnostic nerve/joint blocks) \\
Provocation test & a maneuver that deliberately loads one structure to see whether it reproduces the pain \\
Centralization & low-back pain retreating toward the midline under repeated movement — a good prognostic sign \\
Facet joint / sacroiliac joint (SIJ) / disc & three low-back structures that each produce pain in overlapping regions \\
Dermatome / radiculopathy & the skin area served by one spinal nerve root / pain from an irritated nerve root \\
Modic changes / SPECT & MRI vertebral-marrow signal changes / a nuclear-imaging scan — both raise a source's likelihood \\
Central sensitization / nociplastic pain & a contested state in which the nervous system amplifies pain centrally rather than from a peripheral lesion \\
Quantitative sensory testing (QST) & standardized psychophysical pain measures (temporal summation, conditioned pain modulation, pressure-pain threshold) \\
Ecological momentary assessment (EMA) & repeated real-time symptom sampling, usually by phone \\
Two-point discrimination / tactile acuity & the smallest separation at which two touches feel like two — spatial resolution on the skin \\
High-sensitivity troponin & a blood marker of heart-muscle injury used to diagnose myocardial infarction \\
Hazard ratio & the relative rate of an event (e.g., death) between two groups over time \\
Insula (anterior / posterior) & a brain region; its \emph{posterior} part codes sensory/spatial pain, its \emph{anterior} part affective distress \\
\bottomrule
\end{xltabular}}

\section{Mathematics \& statistics terms (for the clinical and psychology reader)}

{\footnotesize\begin{xltabular}{\textwidth}{>{\raggedright\arraybackslash}X>{\raggedright\arraybackslash}X>{\raggedright\arraybackslash}X}
\toprule
Term & Plain meaning &  \\
\midrule
\endhead
Consilience & agreement between independent lines of evidence that were gathered for unrelated reasons; convergence from different directions counts as stronger support than repetition from one &  \\
Inverse problem & recovering hidden causes from their observed effects (here, the source from the pain map) &  \\
Identifiability & whether the causes can be uniquely recovered from the data at all &  \\
Rank-deficient / null space & a mapping that collapses distinct inputs to the same output; the null space = causes giving an identical observation &  \\
Lead-field matrix & the mapping from internal sources to surface measurements (borrowed from EEG source localization) &  \\
Mutual information (bits) & how much observing one variable reduces uncertainty about another &  \\
Prior / likelihood / posterior & Bayesian ingredients: prior belief, the data's evidence, the updated belief &  \\
Bayesian observer model & a model of a perceiver who combines noisy evidence with prior belief as probability theory prescribes; used as a benchmark for how well a person could possibly do &  \\
Generative model & a probabilistic recipe for how causes produce data; diagnosis inverts it &  \\
Node (of a model) & one factor in the \emph{probabilistic} generative model, and one place it can fail: the likelihood, the model class, group- or context-dependence in that same likelihood, or the report. Decision and loss are a separate downstream layer, not a node of this model (companion §3) &  \\
Repulsion & a report pushed \emph{away} from a cue rather than toward it; the opposite of the pull that cue integration produces, and outside what a convex blend can generate &  \\
Bifurcation / phase transition / critical point & a qualitative regime change at a threshold. Dynamical systems says \emph{bifurcation}, statistical physics says \emph{phase transition}, for the same event &  \\
Neural field & a continuous model of cortical activity over space (Amari, 1977); can hold a stable “bump” or spread &  \\
Controlled direct effect / causal mediation & the effect of one variable holding a mediator fixed at a specified level, under potential-outcomes assumptions — isolates a mechanism. Stronger than the standardized policy contrast this paper actually computes (companion §6) &  \\
Standardized policy contrast (under current uptake) & the outcome change from a detection policy alone, holding treatment \emph{uptake} at its currently observed group-specific rate — an algebraic decomposition, not a controlled direct effect &  \\
Covariate shift & in machine learning, usually means the input distribution $p(X)$ changes while $p(Y \mid X)$ stays fixed &  \\
Group-conditional presentation shift & this paper's term for what mechanism (c) hypothesizes: $p(\text{location} \mid \text{diagnosis}, \text{group}, \text{context})$ itself varies by group — a stronger and different claim than covariate shift, and still open (Table 3) &  \\
Statistical power / partial correlation / Cohen's f$^{2}$ & chance of detecting a true effect / association after adjustment / a standardized effect size &  \\
Funnel plot / Egger test & tools that detect publication bias from asymmetry across study results &  \\
\bottomrule
\end{xltabular}}

\section{Pain-science terms (for the mathematics reader)}

{\footnotesize\begin{xltabular}{\textwidth}{>{\raggedright\arraybackslash}X>{\raggedright\arraybackslash}X}
\toprule
Term & Plain meaning \\
\midrule
\endhead
Referred pain & pain felt at a site distant from its actual source (e.g., cardiac pain in the arm) \\
Anatomical multiplexing \emph{(our term)} & many structures sharing one felt location — mechanism (a) \\
Delocalized amplification \emph{(our term)} & centrally amplified pain that spreads and migrates — mechanism (b) \\
Group-conditional presentation shift \emph{(our term)} & a systematic, person-dependent shift of felt location — mechanism (c) \\
Dorsal-horn convergence & skin, muscle, and visceral inputs feeding the same spinal neurons — the anatomical basis of referral \\
Proximal$\to $distal projection bias & a tendency to feel deep pain more distally, from denser distal cortical representation \\
Body-perception disturbance / mislocalization & a distorted sense of where one's own body or pain is, seen in chronic pain \\
\bottomrule
\end{xltabular}}

\begin{quote}\itshape We spend our days wary of every ache,\\though joy itself is edged with pain.\\Mistaken for an enemy, it keeps its watch,\\the sentinel that wounds in order to protect.\end{quote}
\end{document}